\documentclass[usenatbib]{mn2e}

\usepackage{savesym}
\usepackage{amsmath}
\savesymbol{iint}
\usepackage{txfonts}
\restoresymbol{TXF}{iint}

\usepackage{subfigure}
\usepackage{epsfig}
\usepackage{enumerate}
\usepackage{url}
\def\simless{\mathbin{\lower 3pt\hbox
{$\rlap{\raise 5pt\hbox{$\char'074$}}\mathchar"7218$}}}   
\def\simmore{\mathbin{\lower 3pt\hbox
{$\rlap{\raise 5pt\hbox{$\char'076$}}\mathchar"7218$}}}   
\newcommand{\be}{\begin{equation}}
\newcommand{\ee}{\end{equation}}
\newcommand       \bea          {\begin{eqnarray}}
\newcommand       \eea          {\end{eqnarray}}
\newcommand       \apj          {ApJ}
\newcommand       \apjl         {ApJL}
\newcommand       \aap          {A\&A}
\newcommand       \nat          {Nature}
\newcommand \rmxaa {Revista Mexicana de Astronomia y Astrofisica}
\newcommand       \mnras        {MNRAS}

\newcommand       \aj      {AJ}
\newcommand       \prd      {Phys.~Rev.~D.~}

\newcommand       \araa      {ARA\&A}

\def\simlt{\mathrel{\hbox{\rlap{\hbox{\lower4pt\hbox{$\sim$}}}\hbox{$<$}}}}
\def\simgt{\mathrel{\hbox{\rlap{\hbox{\lower4pt\hbox{$\sim$}}}\hbox{$>$}}}}
\def\lesssim{\mathrel{\hbox{\rlap{\hbox{\lower4pt\hbox{$\sim$}}}\hbox{$<$}}}}

\def\gtrsim{\mathrel{\hbox{\rlap{\hbox{\lower4pt\hbox{$\sim$}}}\hbox{$>$}}}}

\title[Optical Emission from Tidal Disruption Events]{A Bright Year for Tidal Disruptions?}

\author[]{Brian D.~Metzger$\thanks{E-mail: bmetzger@phys.columbia.edu}$, Nicholas C. Stone\\
Columbia Astrophysics Laboratory, Columbia University, New York, NY, 10027, USA\\ 
}
\begin{document}
\date{Received / Accepted}
\pagerange{\pageref{firstpage}--\pageref{lastpage}} \pubyear{2013}

\maketitle

\label{firstpage}

\begin{abstract}
When a star is tidally disrupted by a supermassive black hole (BH), roughly half of its mass falls back to the BH at super-Eddington rates.  Being tenuously gravitationally bound and unable to cool radiatively, only a small fraction $f_{\rm in} \ll 1$ of the returning debris will likely be incorporated into the disk and accrete, with the vast majority instead becoming unbound in an outflow of velocity $\sim 10^{4}$ km s$^{-1}$.  This slow outflow spreads laterally, encasing the BH.  For months or longer, the outflow remains sufficiently neutral to block hard EUV and X-ray radiation from the hot inner disk, which instead becomes trapped in a radiation-dominated nebula.  Ionizing nebular radiation heats the inner edge of the ejecta to temperatures of $T \gtrsim$ few 10$^{4}$ K, converting the emission to optical/near-UV wavelengths where photons more readily escape due to the lower opacity.   This can explain the unexpectedly low and temporally constant effective temperatures of optically-discovered TDE flares.  For BHs with relatively high masses $M_{\bullet} \gtrsim 10^{7}M_{\odot}$ the ejecta can become ionized at an earlier stage, or for a wider range of viewing angles, producing a TDE flare which is instead dominated by thermal X-ray emission.  We predict total radiated energies consistent with those of observed TDE flares, and ejecta velocities that agree with the measured emission line widths.  The peak optical luminosity for $M_{\bullet} \lesssim 10^{6}M_{\odot}$ is suppressed as the result of adiabatic losses in the inner disk wind, possibly contributing to the unexpected dearth of optical TDEs in galaxies with low mass BHs.  In the classical picture, where $f_{\rm in} \approx 1$, TDEs de-spin supermassive BHs and cap their maximum spins well below theoretical accretion physics limits.  This cap is greatly relaxed in our model, and existing Fe K$\alpha$ spin measurements provide suggestive preliminary evidence that $f_{\rm in} < 1$.
\end{abstract} 
  


\section{Introduction}

Occasionally a star in a galactic nucleus is perturbed onto a highly eccentric orbit that brings it close enough to the central supermassive black hole (BH) to be torn apart by tidal forces (\citealt{Hills75}; \citealt{Carter&Luminet83}).  The rapid accretion of stellar debris following such a tidal disruption event (TDE) was predicted to power a luminous electromagnetic flare, detectable to large distances (\citealt{Rees88}).  

TDE flares have been discovered at hard X-ray (e.g., \citealt{Bloom+11}, \citealt{Burrows+11}; \citealt{Cenko+12}, \citealt{Pasham+15}), soft X-ray (e.g., \citealt{Bade+96}, \citealt{Grupe+99}, \citealt{Komossa&Bade99}, \citealt{Komossa&Greiner99}, \citealt{Donley+02}, \citealt{Esquej+08}, \citealt{Maksym+10}, \citealt{Saxton+12}), ultraviolet (e.g., \citealt{Stern+04},  \citealt{Gezari+06,Gezari+08,Gezari+09}), and optical (\citealt{VanVelzen+11}, \citealt{Gezari+12}, \citealt{Cenko+12b}, \citealt{Arcavi+14}, \citealt{Chornock+14}, \citealt{Holoien+14}, \citealt{Vinko+15}) wavelengths (see \citealt{Komossa15} for a recent review).  These events are generally characterized by angular positions consistent with the centers of their host galaxies; luminosities close to or exceeding the Eddington luminosity for the inferred BH mass; and a characteristic light curve decay similar to the predicted $\propto t^{-5/3}$ fall-back rate of stellar debris following a TDE (\citealt{Rees88}; \citealt{Ayal+00}; \citealt{Lodato+09}).  

Despite some similarities, it is not yet clear how this heterogeneous class of events, detected by surveys at different wavelengths with different selection criteria and biases, fits together into a coherent picture of the TDE phenomenon.  For instance, while some jetted TDEs (those characterized by nonthermal X-ray and synchrotron radio emission) show luminous thermal optical emission (\citealt{Pasham+15}), most optical and soft X-ray discovered flares are radio quiet (\citealt{Bower+13}; \citealt{VanVelzen+13}).  The first TDEs were detected in soft X-rays, yet many optically-discovered events produce no detectable X-ray emission (e.g. PS1-10jh, \citealt{Gezari+12}, and ASAS-SN14ae, \citealt{Holoien+14}), or do so only after a significant delay (e.g., D3-13, \citealt{Gezari+06} and D1-9, \citealt{Gezari+09}).

The peak effective temperature of thermal radiation produced by radiatively-efficient accretion onto a BH of mass $M_{\bullet} = 10^{6}M_{\bullet,6}M_{\odot}$ is given by (e.g.,~\citealt{Miller15})
\be
T_{\rm eff} \approx 0.54\left(\frac{\dot{M}}{\dot{M}_{\rm Edd}}\frac{G^{2}M_{\bullet}^{2}}{\kappa_{\rm es}\sigma \eta c R_{\rm in}^{3}}\right)^{1/4} \approx 4\times 10^{5}\,{\rm K\,\,}\left(\frac{\dot{M}}{\dot{M}_{\rm Edd}}\frac{0.1}{\eta}\right)^{1/4}M_{\bullet,6}^{-1/4},
\label{eq:Teffd}
\ee
where $\dot{M}$ is the accretion rate normalized to the Eddington rate of $\dot{M}_{\rm Edd} = 4\pi GM_{\bullet}/(\kappa_{\rm es}\eta c)$, $\kappa_{\rm es} \approx 0.4$ cm$^{2}$ g$^{-1}$ is the electron scattering opacity, $\eta = L_{\rm acc}/\dot{M}c^{2}$ is the radiative efficiency, $L_{\rm acc}$ is the accretion luminosity, and $R_{\rm in} = 6R_{\rm g}$ is the radius of the innermost stable orbit for a non-rotating BH in units of the gravitational radius $R_{\rm g} \equiv GM_{\bullet}/c^{2}$.

One of the puzzles of the optically-discovered TDEs is why their measured color temperatures are consistently much lower than those predicted by standard accretion models.  The best-fit Planck temperatures are usually in the range of $\approx 2-5\times 10^{4}$ K (\citealt{VanVelzen+11}, \citealt{Gezari+12}, \citealt{Arcavi+14}, \citealt{Chornock+14}, \citealt{Holoien+14}, \citealt{Vinko+15}), well below the predicted values of $T_{\rm eff} \approx 3\times 10^{5}-10^{6}$ K from equation (\ref{eq:Teffd}) for $M_{\bullet} \lesssim 10^{7}$ K and the range of accretion rates, $\dot{M} \gtrsim \dot{M}_{\rm Edd}$, thought to characterize TDEs.  PS1-10jh maintained a constant optical-UV spectral slope, even as the total flux decreased by more than a factor of 10 (\citealt{Gezari+12}).  

The low temperatures of TDE flares have generally been attributed to the presence of a ``reprocessing layer," which absorbs higher energy radiation from the inner disk and reradiates it at lower frequencies (\citealt{Loeb&Ulmer97}, \citealt{Guillochon&RamirezRuiz13}).  If such a layer is in thermal equilibrium, then to explain the observed temperatures the layer should be located at radii of $\gtrsim 10^{3}-10^{4} R_{\rm g}$, which are much larger than the radius $R_{\rm c} \lesssim 100 R_{\rm g}$ at which the stellar debris will circularize into the accretion disk.  

One possible source of radially-extended matter are collisions between the eccentric streams of stellar debris created during the disruption, which may contribute to the process of debris circularization (e.g., \citealt{Hayasaki+13}, \citealt{Cheng&Bogdanovic14}, \citealt{Hayasaki+15}, \citealt{Shiokawa+15}, \citealt{Piran+15}, \citealt{Bonnerot+15}).  However, this explanation begs the question: if circularization is so inefficient as to result in such a massive and long-lived reprocessing layer, then why do observed TDE light curves appear to so faithfully track the predicted $\propto t^{-5/3}$ decay in the fall-back rate?  \citet{Miller15} proposed that the low temperatures of TDE flares result from line-driven winds from an extended accretion disk, which suppresses the fraction of the accreting gas which reaches the hot inner disk.  However, this requires a Keplerian disk that extends to thousands of gravitational radii, which the stellar debris in a TDE does not have enough angular momentum to create.  Nevertheless, the idea that the fraction of the infalling gas that ultimately reaches small radii to accrete is reduced may well be correct.  

A defining feature of TDE fall-back is the very low angular momentum of the infalling gas compared to its gravitational binding energy (\citealt{Coughlin&Begelman14}).  Gas inflowing at super-Eddington rates cannot radiatively cool on the dynamical or accretion timescale, in which case photons are trapped and advected with the fluid.  Low initial binding energy coupled with this globally adiabatic evolution suggests that only a fraction of the infalling debris is likely to remain gravitationally bound to the BH, with the remainder unbound in an outflow (\citealt{Strubbe&Quataert09}; \citealt{Lodato&Rossi11}).  

This work develops an alternative model for thermal TDE emission, and the source of the reprocessing material, based on the assumption that the accreted fraction of gas is very low and that most of the disrupted star that begins weakly bound to the BH is promptly ejected soon after reaching pericenter.  Such a massive outflow is necessarily slow and, in contrast to the finding of past work (\citealt{Strubbe&Quataert09}), is found to remain opaque to hard EUV/soft X-ray radiation from the inner disk for an extended period of months or longer.  The ``reprocessing" of high frequency radiation from the inner disk due to selective absorption by this neutral outflow will regulate the effective temperatures of some TDE flares to be $\sim $ few 10$^{4}-10^{5}$ K, helping to explain the low and temporally constant colors of the optically-discovered events.  Our results also provide insight into a potential unification of thermal TDE phenomenon, by delineating the conditions under which X-rays from the inner disk will promptly ionize their way out through the outflow ejecta, resulting instead in a luminous X-ray flare, at least for some range of viewing angles. 

\begin{figure*}
\includegraphics[width=1.0\textwidth]{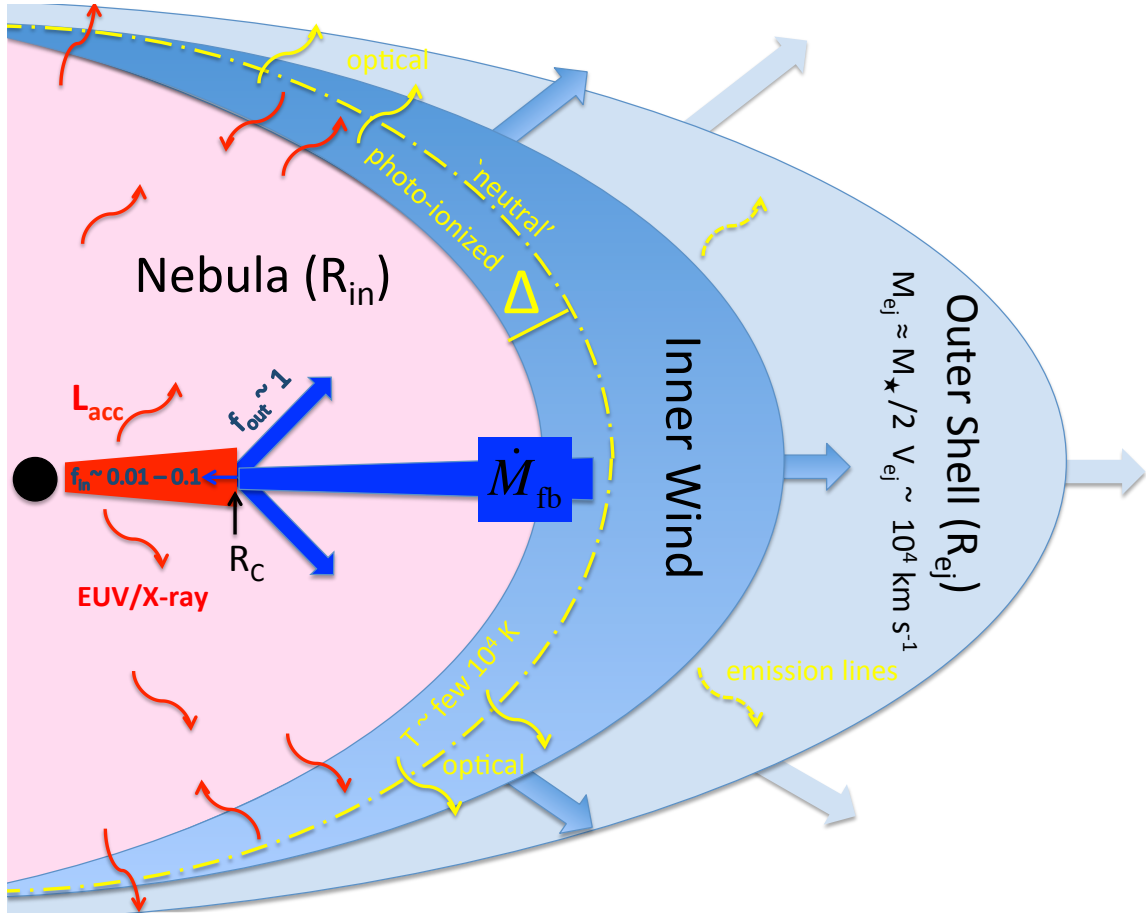}
\caption{Schematic illustration of TDE inflow-outflow geometry and thermal emission.  Stellar debris following the disruption returns to pericenter at a rate $\dot{M}_{\rm fb}$ ({\it dark blue}).  Due to the low gravitational binding energy of the debris and its inability to cool radiatively (eq.~[\ref{eq:Eratio}]), only a small fraction $f_{\rm in} \sim 0.01-0.1$ joins the inner disk and accretes onto the black hole ({\it red}).  The vast majority is instead unbound in a wind ({\it blue}) of velocity $v_{\rm ej} \sim$ few 10$^{3}-10^{4}$ km s$^{-1}$ (eqs.~[\ref{eq:vej1}], [\ref{eq:vejmax}]), which accumulates in an outer shell of radius $R_{\rm ej}$ and mass $M_{\rm ej} \approx M_{\star}/2$ ({\it light blue}).  Extreme EUV/X-ray radiation from the inner accretion disk ({\it red photons}) is initially trapped behind the expanding wind ejecta in a hot nebula of radius $R_{\rm in}$ ({\it pink}).  Nebular radiation ionizes the inner wind to a depth $\Delta$ ({\it dot-dashed yellow line}), which remains less than the ejecta thickness for several fall-back times ($t_{\rm ion} \gtrsim $ few $t_{\rm fb}$; eq.~[\ref{eq:tion}]) for massive BHs ($M_{\bullet} \lesssim 10^{7}M_{\odot}$; Fig.~\ref{fig:timescales}).  This ionized layer is heated to a temperature of $\sim$ few 10$^{4}$ K, similar to other astrophysical environments with similar ionization parameters such as the ejecta in classical novae.  Optical radiation from this layer more easily escapes from the ejecta due to the lower electron scattering ({\it yellow photons}), producing a luminous flare.  Lower densities perpendicular to the plane of disruption may allow X-rays to escape from this nebula preferentially along these polar directions, resulting in a possible viewing angle dependence of the thermal X-ray luminosity.  Broad emission lines of hydrogen or helium ({\it dashed yellow photon}) may originate from the photo-ionized region, or from larger radii in the inner wind or outer shell.} 
\label{fig:schematic}
\end{figure*}

This paper is organized as followed.  In $\S\ref{sec:TDE}$ we briefly review essential elements of TDE physics.  In $\S\ref{sec:outflows}$ we describe the properties of the unbound outflow, and of the smaller fraction which is accreted by the BH.  In $\S\ref{sec:radiation}$ we present a toy model for the evolution of the hard (ionizing) and soft (optical) radiation behind the ejecta.  This model describes the physical process by which the accretion power is converted to optical radiation and quantifies the radiative efficiency of TDEs.  Section $\S\ref{sec:ion}$ describes the ionization state of the ejecta, exploring over what range of conditions hard-UV/X-ray photons are trapped, with implications for the diversity of thermal TDE flare phenomena.  In $\S\ref{sec:temperature}$ we describe the temperature of the reprocessed emission.  In $\S\ref{sec:spin}$ we discuss an implications of our model for the cosmological evolution of the mass and spin of supermassive BHs, and discuss how high measured spin values provide tentative evidence for a low accretion efficiency in TDEs.  In $\S\ref{sec:discussion}$ we discuss our results and summarize our conclusions.  Figure \ref{fig:schematic} provides a schematic illustration of our model, the pieces of which are developed as we proceed.  

\section{Tidal Disruption Events}
\label{sec:TDE}

A star of mass $M_{\star} = m_{\star}M_{\odot}$ and radius $R_{\star}$ is tidally disrupted if the pericenter radius of its orbit, $R_{\rm p}$, becomes less than the tidal radius
\be
R_{\rm t} = R_{\star}(M_{\bullet}/m_{\star})^{1/3} \approx 7\times 10^{12}\,{\rm cm} \,m_{\star}^{7/15}M_{\bullet,6}^{1/3} \approx 47 m_{\star}^{7/15}M_{\bullet,6}^{-2/3}R_{\rm g},
\label{eq:Rt}
\ee
where we have assumed a mass-radius relationship $R_{\star} \approx m_{\star}^{4/5}R_{\odot}$ appropriate to lower main sequence stars (\citealt{Kippenhahn&Weigert90}).  The orbital penetration factor is defined as $\beta \equiv R_{\rm t}/R_{\rm p} > 1$, with a maximum value allowed for disruption outside the horizon of
\be
\beta_{\rm max} \approx 12 m_{\star}^{7/15}M_{\bullet,6}^{-2/3}(R_{\rm ibco}/4R_{\rm g})^{-1},
\label{eq:betamax}
\ee
where $R_{\rm ibco}$ is the radius of the innermost bound circular orbit.  The maximum BH mass ($\beta_{\rm max} = 1$) resulting in a disruption is thus 
\be
M_{\rm \bullet,max} \approx 4\times 10^{7}M_{\odot} m_{\star}^{7/10}(R_{\rm ibco}/4R_{\rm g})^{-3/2},
\ee
which depends on the BH spin $a$ through $R_{\rm ibco}(a)$, where, e.g., $R_{\rm ibco} = 4R_{\rm g}$ for a non-spinning BH.  This approximate treatment of the highly relativistic regime near the horizon is in reasonable agreement with the relativistic hydrodynamics models of \citet{Ivanov&Chernyakova06}.

Disruption binds roughly half the star to the BH by a specific energy $|E_{\rm t}| \sim GM_{\bullet}R_{\star}/R_{\rm t}^{2}$ corresponding roughly to the work done by tidal forces over a distance $\sim R_{\rm t}$.  The most tightly bound matter falls back to the BH on the characteristic fall-back timescale set by the period of an orbit with energy $E_{\rm t}$,
\be
 t_{\rm fb} \simeq 41\,{\rm d}\,M_{\bullet,6}^{1/2}m_{\star}^{1/5}
\label{eq:tfb}
\ee
which is generally, to first order, independent of the penetration factor $\beta$ (\citealt{Stone+13}; \citealt{Guillochon&RamirezRuiz13}).  The resulting rate of mass fall-back at times $t \gg t_{\rm fb}$ is given by
\be
\dot{M}_{\rm fb} \approx \frac{M_{\star}}{3 t_{\rm fb}}\left(\frac{t}{t_{\rm fb}}\right)^{-5/3} \approx 2\times 10^{26}\,\,{\rm g\,s^{-1}}\,M_{\bullet,6}^{-1/2}m_{\star}^{4/5}\left(\frac{t}{t_{\rm fb}}\right)^{-5/3}.
\label{eq:Mdotfb}
\ee
The peak Eddington ratio of the fall-back rate is given by
\be
\frac{\dot{M}_{\rm fb}(t_{\rm fb})}{\dot{M}_{\rm Edd}} \approx 133 \eta_{-1}M_{\bullet,6}^{-3/2}m_{\star}^{4/5},
\label{eq:edd}
\ee
where $\eta = 0.1\eta_{-1}$.  
  

\section{Inflows and Outflows}
\label{sec:outflows}

Stellar debris on a nearly parabolic orbit returns to the BH with an angular momentum sufficient to circularize at a radius $R_{\rm c} \simeq 2R_{\rm p} = 2R_{\rm t}\beta^{-1}$.  The poorly understood process by which matter dissipates its orbital energy is an area of active research (\citealt{Kochanek94}; \citealt{Guillochon&RamirezRuiz13}; \citealt{Hayasaki+15}; \citealt{Shiokawa+15}).  In this work we assume that circularization of at least a fraction of the debris is efficient on timescales $\sim t_{\rm fb}$.  We further assume that the narrow infalling tidal streams are sufficiently dense to reach the disk by penetrating the outflowing matter distributed over a larger solid angle (Fig.~\ref{fig:schematic}).  

We assume also that only a small fraction $f_{\rm in} \ll 1$ of the gas that returns to pericenter becomes part of the Keplerian disk.  Gas infalling at rates well above the Eddington rate (eq.~[\ref{eq:edd}]) cannot radiatively cool on the dynamical time (e.g., \citealt{Strubbe&Quataert09}).  Quantitatively, the timescale for radiative cooling of gas inflowing with density $\rho \approx \dot{M}_{\rm fb}/(4\pi v_{\rm ff}R_{\rm p}^{2})$ near pericenter can be estimated as the diffusion timescale $t_{\rm d} \sim \tau_{\rm es}(R_{\rm p}/c) \approx \dot{M}_{\rm fb}\kappa_{\rm es}t_{\rm dyn}/(4\pi c R_{\rm p})$, where $v_{\rm ff} \sim R_{\rm p}/t_{\rm dyn}$ is the free-fall velocity and $\tau_{\rm es} \sim \rho R_{\rm p}\kappa_{\rm es}$ is the electron scattering optical depth near pericenter.  The ratio of the cooling timescale to the dynamical timescale,
\be
\frac{t_{\rm d}}{t_{\rm dyn}} \sim \frac{\dot{M}_{\rm fb} \kappa_{\rm es}}{4\pi c R_{\rm p}} \approx 10\beta \frac{\dot{M}}{\dot{M}_{\rm Edd}}\left(\frac{R_{\rm g}}{R_{\rm t}}\right) \approx 28\, \beta m_{\star}^{1/3}M_{\bullet,6}^{-5/6}\left(\frac{t}{t_{\rm fb}}\right)^{-5/3},
\label{eq:tdperi}
\ee
thus exceeds unity until the ``cooling" time
\begin{eqnarray}
t_{\rm cool}/t_{\rm fb} \sim 7.4 \beta^{3/5}m_{\star}^{1/5}M_{\bullet,6}^{-1/2} \gg 1,
\label{eq:trad}
\end{eqnarray}
where we have used equations (\ref{eq:Rt}) and (\ref{eq:Mdotfb}).  Note that equation (\ref{eq:trad}) is a lower limit on the cooling time, since $t_{\rm d}$ in equation (\ref{eq:tdperi}) is derived assuming the density of a spherical inflow and the inflowing tidal debris will be much denser.  

At times $t \ll t_{\rm cool}$, gas returning to pericenter is effectively adiabatic on the dynamical timescale.\footnote{Comparing the cooling and dynamical times near pericenter is not the relevant criterion for energy released during the circularization process at larger radii, e.g. near stream apocenter.  However, since the total energy dissipated in the circularization process is $\propto 1/r$ and hence will be dominated by the energy released on the smallest radial scale $\sim R_{\rm p}$, equation (\ref{eq:tdperi}) eventually becomes the relevant criterion.}  On the other hand, the specific energy that must be removed to circularize the debris, $|E_{\rm c}| = GM_{\bullet}/2R_{\rm p}$, is typically\footnote{Speculatively, the value of $E_{\rm t}$ could be dramatically increased for rare deeply penetrating events; at high $\beta$, the interaction between tidal compression and misaligned BH spin may be capable of binding a large fraction of the debris much more tightly to the BH \citep{Stone+13}, but this has not been simulated numerically.  Recent numerical results indicate such an enhancement may be possible for extreme $\beta$ even absent BH spin \citep{Evans+15}, although such an enhancement can also arise spuriously in hydrodynamical simulations of TDEs that under-resolve the orbital midplane \citep{Guillochon+09}.} two orders of magnitude larger than the orbital binding energy of the debris $|E_{\rm t}| \sim GM_{\bullet}R_{\star}/R_{\rm t}^{2}$,
\be
E_{\rm t}/E_{\rm c} \sim R_{\star}\beta^{-1}/R_{\rm t} \approx 0.01\beta^{-1}m_{\star}^{1/3}M_{\bullet,6}^{-1/3}.
\label{eq:Eratio}
\ee

Hydrodynamical simulations by \citet{Li+13} explore such tenuously gravitationally bound (``zero energy"), radiatively inefficient accretion flows in the context of low luminosity AGN.  They find that only a fraction $f_{\rm in} \sim \alpha$ of the inflowing matter becomes bound to the disk, where $\alpha \approx 0.01-0.1$ is the Shakura Sunyaev effective viscosity parameter.   This finding can be understood as a comparison between the accretion (or viscous) time for a thick disk, $t_{\rm visc} \sim (\Omega_{\rm K}\alpha)^{-1}$, and the time the tidal streams spend near pericenter $t_{\rm orb} \sim \Omega_{\rm K}^{-1}$, where $\Omega_{\rm K}$ is the Keplerian angular frequency.  The remaining fraction $1-f_{\rm in} \approx 1$ is ejected in an outflow from near the circularization radius (see below). 

There are differences between the TDE problem and the set-up of \citet{Li+13}, who evaluate a rotating flow that becomes spherical at large radii and who include no radiation transport.  Nevertheless, a small value of $f_{\rm in} \sim 0.01-0.1$ appears natural when one combines the requirements of (1) energy conservation; (2) globally adiabatic evolution, due to inefficient radiative cooling on timescales short enough to explain observed TDE light curves; (3) matching the measured velocity line widths in TDE flares (see below).  

As an explicit example, consider an alternative picture to the `prompt ejection' scenario outlined above based on \citet{Li+13}.   Imagine that gas does first circularize into a circular rotating configuration without becoming unbound on the dynamical timescale.  The very low binding energy of the circularized matter implies that it will form a geometrically thick torus with substantial radial pressure support (e.g., \citealt{Coughlin&Begelman14}).  If a fraction $\eta$ of the accreted rest mass is released on the accretion timescale $\sim t_{\rm dyn}/\alpha$, the total energy released will be sufficient to unbind the rest of the gas on a timescale $t_{\rm out} \sim E_{\rm b}t_{\rm dyn}/\eta c^{2}$, where the specific binding energy $E_{\rm b} =$ max$[|E_{\rm t}|, |E_{\rm orb}|]$ is the maximum of that due the tidal disruption process, $E_{\rm t}$, and that of the disrupted star's orbit, $E_{\rm orb}$.  In this case, the condition for non-radiative evolution on the timescale $\sim t_{\rm out}$ is that the ratio
\be
\frac{t_{\rm d}}{t_{\rm out}} = \frac{\eta \alpha c^{2}}{E_{\rm b}}\frac{t_{\rm d}}{t_{\rm dyn}} \approx 0.48\eta_{-1} m_{\star}^{22/15}M_{\bullet,6}^{-5/3}\left(\frac{\alpha}{0.1}\right)\frac{t_{\rm d}}{t_{\rm dyn}}
\label{eq:trad2}
\ee
exceeds unity, where in the final equality we have assumed that $E_{\rm b} = E_{\rm t}$, i.e. that $|E_{\rm t}| \gg |E_{\rm orb}|$, as is usually well satisfied.  For $\alpha \sim 0.1$, the condition $t_{\rm d} > t_{\rm out}$ is similar to the $t_{\rm d} > t_{\rm exp}$ condition given above in the prompt ejection scenario (eq.~[\ref{eq:tdperi}]) for characteristic parameters, e.g. $M_{\bullet,6}, \eta_{-1}, m_{\star} \sim 1$.  Thus, according to either criterion, the debris cannot efficiently cool for several fall-back times.  This renders outflows likely, except in the unlikely occurrence that the majority of the excess thermal energy is advected across the BH horizon (\citealt{Narayan&Yi95}). 

Our assumption that a large fraction of the inflowing tidal debris is promptly unbound appears to be incompatible with the findings of recent hydrodynamic simulations of debris circularization in TDEs, which do not
show such mass-loaded outflows despite their use of adiabatic equations of state (\citealt{Hayasaki+15}, \citealt{Shiokawa+15}, \citealt{Bonnerot+15}).  These simulations do not, however, include magnetic fields, which may be responsible for the angular momentum transport (viscosity) that could preferentially bind some mass at the expense of unbinding the rest.  Perhaps more importantly, for numerical reasons some of these works simulate the disruption of stars on unrealistically low eccentricity orbits, for which the initial gravitational binding energy of the star is much larger than the more physical case of a nearly parabolic (zero energy) orbit.  This is
true for all the simulations of \citet{Hayasaki+15}, and some of \citet{Bonnerot+15}, where eccentricity $e \approx 0.8$ and $E_{\rm b}/E_{\rm c} \sim 0.1$.  This artificially high ratio renders these simulations at best
marginally capable of depositing sufficient energy to drive an outflow from scales $R \approx R_{\rm p}$\footnote{\citet{Bonnerot+15} also simulate one TDE with $e=0.95$ and a more physical $E_{\rm b}/E_{\rm c} \sim 0.01$, but run this simulation for much less than a disk viscous time, which (for
$\alpha \sim 0.01$) would be needed to launch the outflow.}.  \citet{Shiokawa+15} initializes a parabolic orbit, but they employ an artificially small BH to star mass ratio of 500:1, which also results in a large initial
binding energy of $E_{\rm t}/E_{\rm c} \sim 0.1$, according to equation (\ref{eq:Eratio}).  Although enough energy is released in shock dissipation to in principle unbind some gas from the BH, this energy is quickly advected inward through an artificially large inner boundary (\citet{Shiokawa+15}, their Fig. 14).

Even if these hydrodynamic simulations were run for more physical mass ratios and initial orbital eccentricities, the issue of outflows might still not be satisfactorily resolved.  The importance of unbound outflows in radiatively inefficient accretion flows is an issue of long-standing debate, even in cases when the disk begins tightly gravitationally bound near its outer edge (\citealt{Stone+99}, \citealt{Blandford&Begelman99}; \citealt{Quataert&Gruzinov00}; \citealt{Hawley&Balbus02}; see \citealt{Yuan&Narayan14} for a recent review).  Global MHD simulations (e.g., \citealt{Hawley&Balbus02}) find that a large fraction of the disk mass is lost to disk outflows within a decade of radius of its outer disk, with a velocity close to the escape speed from this location.  TDE fall-back debris is even more weakly bound (lower Bernoulli parameter) than the initial conditions used in these simulations.
 
The small fraction of $\dot{M}_{\rm fb}$ which is added to the inner Keplerian disk feeds the BH, producing an accretion luminosity
\be
L_{\rm acc} = \eta f_{\rm in}\dot{M}_{\rm fb}c^{2} \approx 2\times 10^{44}\,{\rm erg\,s^{-1}}\,\eta_{-1}f_{\rm in,-2}M_{\bullet,6}^{-1/2}m_{\star}^{4/5}(t/t_{\rm fb})^{-5/3},
\label{eq:Ld}
\ee
where $f_{\rm in,-2} \equiv f_{\rm in}/0.01$.  The radiative efficiency of geometrically thin accretion varies from $\eta \approx 0.04-0.42$, depending on the spin of the BH and its orientation relative to the angular momentum of the accreting gas.  Remarkably, a relatively high efficiency of $\eta \approx 0.05$ for a non-rotating BH was found by radiation (magneto-)hydrodynamical simulations of highly super-Eddington disk accretion (e.g.~\citealt{Jiang+14}; see also \citealt{Ohsuga+05}; \citealt{Sadowski&Narayan15}; \citealt{Jiao+15}), when both radiative and kinetic luminosity are considered.  This is contrary to the expectations of previous analytic models (e.g.~\citealt{Shakura&Sunyaev73}).  In our model as well, $\eta$ is an agnostic parameter accounting for both radiation and the accretion power placed into the kinetic energy of fast winds or jets from the inner disk.  Such outflows will rethermalize after colliding with the inner layers of the bulk ejecta, in effect producing hard radiation$-$albeit indirectly.   By assuming that the BH accretion rate directly tracks the fall-back rate, we are also neglecting viscous spreading of the accretion disk, which can become relevant on longer timescales (\citealt{Cannizzo+90}; \citealt{Shen&Matzner14}).

The majority of the stellar mass is instead lost in an outflow from near pericenter at a rate of $\dot{M}_{\rm out} = (1-f_{\rm in})\dot{M}_{\rm fb} \approx \dot{M}_{\rm fb}$, resulting in a total ejecta mass of $M_{\rm ej} \approx M_{\star}/2$.  The kinetic power of the unbound debris, $\dot{M}_{\rm out}v_{\rm ej}^{2}/2$, is that required to bind the smaller accreted fraction $f_{\rm in}$ into the Keplerian disk, $|-GM_{\bullet}f_{\rm in}\dot{M}_{\rm fb}/2R_{\rm c}|$, again under the assumption that the original energy of the debris is effectively zero (eq.~[\ref{eq:Eratio}]).  This results in a minimum asymptotic ejecta velocity of
\be
v_{\rm ej}^{\rm min} = \left(\frac{GM_{\bullet}f_{\rm in}}{(1-f_{\rm in})R_{\rm c}}\right)^{1/2} \underset{f_{\rm in}\ll 1}\approx 3100 \,{\rm km\,s^{-1}}\,f_{\rm in,-2}^{1/2}\beta^{1/2} M_{\bullet,6}^{1/3} m_{\star}^{-7/30}.
\label{eq:vej1}
\ee

For relatively high mass BHs, $v_{\rm ej}^{\rm min}$ will approximately equal the mean velocity of the bulk of the ejecta, and so we take $v_{\rm ej} = v_{\rm ej}^{\rm min}$ in most estimates to follow.  However, at early times accretion energy deposited behind the ejecta is trapped on the expansion timescale ($\S\ref{sec:radiation}$).  Work done on the ejecta by the nebular pressure will in this case convert the accretion luminosity $L_{\rm acc}$ (eq.~[\ref{eq:Ld}]) into outflow kinetic energy, $\dot{M}_{\rm out}v_{\rm ej}^{2}/2$, with high efficiency.  The matter ejected at the early times will thus be accelerated to a maximum velocity of
\be
v_{\rm ej}^{\rm max} \approx \sqrt{2\eta f_{\rm in}}c \approx 1.3\times 10^{4}\,{\rm  km\,s^{-1}}\,\eta_{-1}^{1/2}f_{\rm in,-2}^{1/2},
\label{eq:vejmax}
\ee
which typically exceeds $v_{\rm ej}^{\rm min}$ (eq.~[\ref{eq:vej1}]) by a factor of a few.  The observed line widths in the optical spectra of TDE flares correspond to expansion velocities $\sim 3\times 10^{3}-10^{4}$ km s$^{-1}$ (see Table 7 of \citealt{Arcavi+14} for a compilation), motivating values of $f_{\rm in} \sim 0.01-0.1$ and $\eta \sim 0.1$, consistent with the fiducial values adopted above.  

The outer radius of the ejecta grows with time as
\be R_{\rm ej} = v_{\rm ej}t = 1.1\times 10^{15}\,{\rm cm}\,f_{\rm in,-2}^{1/2}\beta^{1/2}m_{\star}^{-1/30}M_{\bullet,6}^{5/6}(t/t_{\rm fb}),
\label{eq:Rej}
\ee
where we have used equation (\ref{eq:vej1}).  Note that $R_{\rm ej}$ exceeds by several orders of magnitude the outflow launching radius $R_{\rm c}$ (eq.~[\ref{eq:Rt}]) on the fall-back time.  On timescales $t \sim t_{\rm fb}$ over which most of the mass and energy are released, the outflow has plenty of time to spread laterally and encase the BH.   

The geometry of the wind ejecta will be complex, requiring more detailed numerical work to properly characterize in detail.  However, for purposes of a simple model it is natural to decompose the outflow into two components: (1) a quasi-spherical ``outer shell" comprised of most of the total mass, released mainly over the first few fall-back times; and (2) a quasi-steady-state ``inner wind'' comprised of matter unbound within the most recent decade of time since disruption.  We assume that the inner wind becomes quasi-spherical and encases the central EUV/X-ray source by a radius $R_{\rm in} \sim 3R_{\rm c} \sim 6R_{\rm p} $, the precise value of which is also uncertain and will require future simulations to more accurately determine.  A spherical geometry may not accurately characterize the disk outflow if it preferentially emerges in one direction, such as might be the case if the returning debris becomes unbound during the first pericenter passage before ``sampling" the full azimuthal extent of the inner disk.  In $\S\ref{sec:discussion}$ we address how the physical picture changes if the outflow is confined to a solid angle $\ll 4\pi$.    

The outer shell can at times $t \gtrsim$ few $t_{\rm fb}$ be approximated as a homologously expanding shell of mass $\approx M_{\star}/2$, radius $R_{\rm ej}$, thickness $\sim R_{\rm ej}$, and mean density
\be
\rho_{\rm sh} \approx \frac{M_{\star}/2}{4\pi R_{\rm ej}^{3}/3} \approx 2\times 10^{-13}\,{\rm g\,cm^{-3}}\,f_{\rm in,-2}^{-3/2}\beta^{-3/2}M_{\bullet,6}^{-5/2}m_{\star}^{11/10}\left(\frac{t}{t_{\rm fb}}\right)^{-3},
\label{eq:rhoej}
\ee
where we have used equation (\ref{eq:Rej}) for $R_{\rm ej}$.  The electron scattering optical depth is 
\be
\tau_{\rm es,sh} \approx \rho_{\rm sh}\kappa_{\rm es}R_{\rm ej} \approx 80f_{\rm in,-2}^{-1}\beta^{-1}M_{\bullet,6}^{-5/3}m_{\star}^{16/15}(t/t_{\rm fb})^{-2}.
\label{eq:taues}
\ee
The shell will remain optically thick ($\tau_{\rm es,sh} > 1$) until a time
\be
t_{\rm thin}/t_{\rm fb} \approx 8.9 f_{\rm in,-2}^{-1/2}\beta^{-1/2}M_{\bullet,6}^{-5/6} m_{\star}^{8/15}.
\label{eq:tthin}
\ee

The density profile of the steady-state inner wind is given by
\begin{eqnarray}
\rho_{\rm w} &=& \frac{\dot{M}_{\rm fb}(1-f_{\rm in})}{4\pi r^{2}v_{\rm ej}} \nonumber \\
&\underset{f_{\rm in} \ll 1}\approx& 3.0\times 10^{-11}\,{\rm g\,cm^{-3}}f_{\rm in,-2}^{-1/2}\beta^{3/2}M_{\bullet,6}^{-3/2}m_{\star}^{0.1}R_{\rm in,6}^{-2}\left(\frac{r}{R_{\rm in}}\right)^{-2}\left(\frac{t}{t_{\rm fb}}\right)^{-5/3},
\label{eq:rhow}
\end{eqnarray}
and the electron scattering optical depth is given by
\begin{eqnarray}
\tau_{\rm es, w} &=& \int_{\rm R_{\rm in}}^{\rm v_{\rm ej}t} \rho_{\rm w}\kappa_{\rm es}dr \approx \frac{\dot{M}_{\rm fb}\kappa_{\rm es}}{4\pi R_{\rm in}v_{\rm ej}} \nonumber \\
&\approx& 480 f_{\rm in,-2}^{-1/2}\beta^{1/2}M_{\bullet,6}^{-7/6}m_{\star}^{17/30}R_{\rm in,6}^{-1}\left(\frac{t}{t_{\rm fb}}\right)^{-5/3},
\label{eq:tauw}
\end{eqnarray} 
where $R_{\rm in,6} \equiv R_{\rm in}/(6R_{\rm p}$) and we have approximated the upper limit of integration as $\sim v_{\rm ej}t \gg R_{\rm in}$ since the value of $\tau_{\rm es,w}$ is not sensitive to this choice. 

Note that because $\tau_{\rm es,w} \gtrsim \tau_{\rm es,sh}$, the outer shell will become optically thin to Thomson scattering earlier than the inner wind.  The outer shell contains most of the mass, but the inner wind dominates the total column.  This will remain true until the wind shuts off, as may occur once radiative cooling of gas streams inflowing to the disk becomes effective at $t \gtrsim t_{\rm cool}$ (eq.~[\ref{eq:trad}]) and inflowing matter settles into the central disk with higher efficiency.


\section{Optical Light Curve Model}
\label{sec:radiation}

A complete model for the coupled evolution of the nebula and the ejecta would require non-LTE, multi-frequency, multi-dimensional radiation hydrodynamics simulation, which is well beyond the scope of this paper.  Here we instead provide a 3-zone toy model that captures the basic physical processes and provides a model for the optical light curve.

The energy in radiation greatly dominates that of matter in both the ejecta and nebula.  We divide this total radiation energy, $E$, into a ``hard" component, $E_{\rm H}$, and a ``soft" (optical) component, $E_{\rm opt}$.  Hard radiation corresponds to EUV/X-ray photons produced by the inner BH accretion disk at a characteristic temperature $T_{\rm eff} \sim 10^{5}-10^{6}$ K (eq.~[\ref{eq:Teffd}]), forming the nebular region around the BH.  We assume that the soft component is produced exclusively by reprocessing of hard radiation after being absorbed by the ejecta walls.  In $\S\ref{sec:ion}$ we show that at early times the ejecta is sufficiently neutral to be opaque to the hard radiation, such that the hard radiation is absorbed in a thin layer separating the nebula from the bulk of the ejecta.  This ionized layer is heated to temperatures $\sim$ few 10$^{4}$ K, the radiation from which sources the optical emission ($\S\ref{sec:temperature}$).  It turns out that the inner wind is more challenging to ionize (requires a higher ionization parameter) than the outer shell ($\S\ref{sec:ion}$).  Hard radiation is thus either confined to the nebula at radii $\gtrsim R_{\rm in}$, or it penetrates the entire ejecta and escapes from the system.  

It is assumed in this section that hard radiation is trapped behind the ejecta, in which case its energy evolves as
\be
\frac{dE_{\rm H}}{dt} = L_{\rm acc} - \frac{E_{\rm H}(1-A)}{t_{\rm lc}},
\label{eq:dEHdt}
\ee
where $L_{\rm acc} \propto f_{\rm in}\dot{M}_{\rm fb}$ (eq.~[\ref{eq:Ld}]) accounts for photon or kinetic luminosity from the hot inner disk.  The second term accounts for losses due to the absorption of hard radiation by the neutral ejecta, where $t_{\rm lc} = R_{\rm n}/c$ is the light crossing time of the nebula (the time between photon-ejecta interactions) of radius $R_{\rm n} \lesssim R_{\rm ej}$ and $(1-A)$ is the probability of absorption per interaction, where $A$ is a frequency-averaged albedo.  \cite{Metzger+14} show using Monte Carlo results (their Fig.~3) that $A \sim 0.4-0.8$ for ratios of the absorptive to scattering opacity, $\zeta \sim 0.01-0.1$, achieved in the ionized layer (see eq.~[\ref{eq:zeta}] below).  Hard radiation is therefore absorbed on at most a few light crossing times, well before adiabatic losses become important.  The steady state solution to equation (\ref{eq:dEHdt}),
\be
E_{\rm H} \approx L_{\rm acc}t_{\rm lc}/(1-A),
\label{eq:EHapprox}
\ee
is thus well satisfied at all times.  Equation (\ref{eq:EHapprox}) will be useful later to estimate under what conditions the ejecta is photo-ionized by the nebular radiation.

The energy in the the soft radiation evolves according to 
\be
\frac{dE_{\rm opt}}{dt} = \frac{E_{\rm H}(1-A)}{t_{\rm lc}}  - \left(\frac{dE_{\rm opt}}{dt}\right)_{\rm ad} - L_{\rm rad},
\label{eq:dEoptdt}
\ee
where the first term accounts for the absorbed hard radiation re-radiated as optical/near-UV radiation (the sink term in eq.~[\ref{eq:dEHdt}]) and the second term accounts for adiabatic losses.  Adiabatic losses can result from both the inner wind and the outer shell.  The outer ejecta shell loses energy due to PdV work at a rate $(dE_{\rm opt}/dt)_{\rm ad} = -(E_{\rm opt}/R_{\rm ej})v_{\rm ej}$ due to the overall expansion of the system.  

The adiabatic loss term for the inner wind is more complicated.  Optical radiation injected at the inner edge of the wind ($r \sim R_{\rm in}$) is advected outwards (``trapped") until it reaches a critical trapping radius $R_{\rm tr} = \dot{M}_{\rm fb}(1-f_{\rm in})\kappa_{\rm es}/(4\pi c)$, defined as the point exterior to which the photon diffusion time is shorter than the outflow expansion time, i.e. $t_{\rm d}/t_{\rm exp} \approx \tau_{\rm es,w}v_{\rm ej}/c < 1$, where $\tau_{\rm es,w} \approx \int_{r} \rho_{\rm w}\kappa_{\rm es}dr \approx \dot{M}_{\rm fb}\kappa_{\rm es}/(4\pi r v_{\rm ej})$ is the Thomson optical depth in the wind exterior to radius $r$ (eq.~[\ref{eq:tauw}]).  

Radiation advected from $r = R_{\rm in}$ to $R_{\rm tr}$ has its temperature decreased adiabatically as $T \propto \rho^{1/3}$, such that the luminosity that escapes is suppressed from its injection value $L_{\rm acc}$ by a factor $\propto T^{4}r^{2} \propto \rho^{4/3}r^{2} \propto r^{-2/3}$ (\citealt{Strubbe&Quataert09}).  In other words, one can define a radiative efficiency
\begin{eqnarray}
f_{\rm rad} \equiv \frac{L_{\rm w}}{L_{\rm acc}} \approx {\rm min}\left[1,\left(\frac{R_{\rm tr}}{R_{\rm in}}\right)^{-2/3}\right],
\label{eq:frad}
\end{eqnarray}
where $L_{\rm w}$ is the luminosity escaping from the inner wind to the outer shell and
\begin{eqnarray}
\left(\frac{R_{\rm tr}}{R_{\rm in}}\right)^{-2/3} \approx \left(\frac{\dot{M}_{\rm fb}\kappa_{\rm es}}{4\pi c R_{\rm in}}\right)^{-2/3} \approx 0.34 \beta^{-2/3} M_{\bullet,6}^{5/9}m_{\star}^{-2/9}R_{\rm in,6}^{2/3}\left(\frac{t}{t_{\rm fb}}\right)^{10/9}.
\label{eq:tr}
\end{eqnarray}
Radiation trapping by the inner wind ($f_{\rm rad} < 1$) will thus suppress the photon luminosity reaching the outer ejecta prior to the time
\be
t_{\rm tr}/t_{\rm fb} \approx 2.6\beta^{3/5}M_{\bullet,6}^{-1/2}m_{\star}^{1/5}R_{\rm in,6}^{-3/5}.
\label{eq:ttr}
\ee

Combining the above, equation (\ref{eq:dEoptdt}) {\it in effect} reads 
\be
\frac{dE_{\rm opt}}{dt} \approx f_{\rm rad}L_{\rm acc}-\frac{E_{\rm opt}}{R_{\rm ej}}v_{\rm ej} - L_{\rm rad},
\label{eq:dEoptdt2}
\ee
where the final term,
\be
L_{\rm rad} = E_{\rm opt}/t_{\rm d},
\label{eq:Lrad}
\ee
accounts for radiative diffusion through the outer shell and represents the observed optical luminosity, where $t_{\rm d} \approx \tau_{\rm es,sh}(R_{\rm ej}/c)$ is the photon diffusion timescale through the outer shell.  No corresponding diffusion term exists in equation (\ref{eq:dEHdt}) for the hard radiation since the opacity to hard radiation is effectively infinite until the ejecta becomes sufficiently photo-ionized ($\S\ref{sec:ion}$). 


Finally, the mean ejecta velocity of the outer shell evolves according to momentum conservation\footnote{We neglect the BH gravitational force, which is unimportant on radial scales $R_{\rm ej} \gg R_{\rm c}$ of interest at times $t \gtrsim t_{\rm fb}$.}
\be
\frac{d}{dt}(M_{\rm ej}v_{\rm ej}) = (1-f_{\rm in})\dot{M}_{\rm fb}v_{\rm w},
\label{eq:momentum}
\ee
where $M_{\rm ej} = \int_0^{t}(1-f_{\rm in})\dot{M}_{\rm fb}(t')dt'$ is the ejecta mass at time $t$.  The source term 
\be
v_{\rm w} = f_{\rm rad}v_{\rm ej}^{\rm min} + (1-f_{\rm rad})v_{\rm ej}^{\rm max}
\ee
 is the instantaneous velocity from the inner wind, which is smoothly interpolated between the earliest phases when the inner wind traps the radiation $f_{\rm rad} \ll 1$ and $v_{\rm w} \approx v_{\rm ej}^{\rm max}$ (eq.~[\ref{eq:vejmax}]) and accretion power primarily goes into accelerating the outflow, and later phases when the inner ejecta is diffusive to radiation ($f_{\rm rad} \sim 1$) and the outflow retains its initial kinetic energy $v_{\rm w} \approx v_{\rm ej}^{\rm min}$ (eq.~[\ref{eq:vej1}]).  The mean asymptotic velocity of the ejecta will always lie between $v_{\rm ej}^{\rm min}$ and $v_{\rm ej}^{\rm max}$ (eq.~[\ref{eq:vejmax}]).  

\subsection{Results}

\begin{figure}
\includegraphics[width=0.5\textwidth]{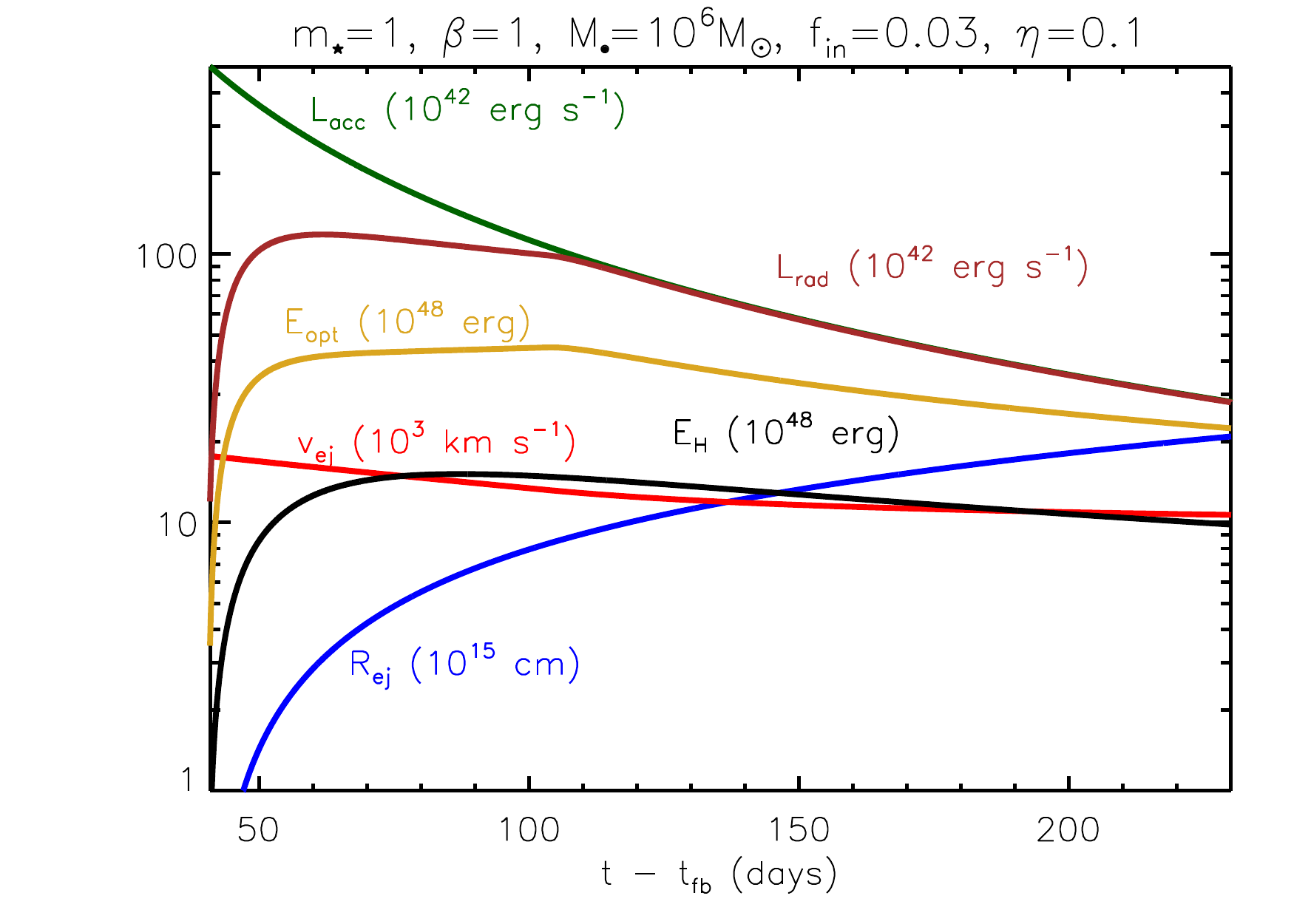}
\caption{Model solution for TDE outflows and optical radiation as a function of time since the fall-back time of the mostly tightly bound debris, calculated for a model with BH mass $M_{\bullet} = 10^{6}M_{\odot}$, stellar mass $M_{\star} = 1M_{\odot}$, penetration factor $\beta = 1$, BH accreted fraction $f_{\rm in} = 0.03$, ejecta albedo $A = 0.5$, inner wind radius $R_{\rm in} = 6R_{\rm p}$, and inner disk radiative efficiency $\eta = 0.1$.  Quantities shown include the accretion power input from the BH, $L_{\rm acc}$  ({\it green}; eq.~[\ref{eq:Ld}]), the radiated optical luminosity, $L_{\rm rad}$ ({\it brown}), mean velocity of outer shell, $v_{\rm ej}$ ({\it red}), outer shell radius $R_{\rm ej}$ ({\it blue}), energy of EUV/X-ray radiation in the nebula, $E_{\rm H}$ ({\it black}), and the energy of optical radiation in the nebula, $E_{\rm opt}$ ({\it gold}). } 
\label{fig:fig1}
\end{figure}

Figure \ref{fig:fig1} shows an example solution of equations (\ref{eq:dEoptdt2}) and (\ref{eq:momentum}) for fiducial parameters: stellar mass $M_{\star} = 1M_{\odot}$, penetration factor $\beta = 1$, BH mass $M_{\bullet} = 10^{6}M_{\odot}$, accreted fraction $f_{\rm in} = 0.03$, accretion radiative efficiency $\eta = 0.1$, and ejecta albedo $A = 0.5$.  At early times the radiated luminosity (brown line) is less than the accretion power of the BH (green line) as the result of adiabatic losses in the inner wind ($f_{\rm rad} < 1$; eq.~[\ref{eq:frad}]).  Within a few fall-back times, however, $L_{\rm rad}$ approaches $L_{\rm acc}$ to high accuracy.  As will be discussed below, this is because once radiation escapes the inner wind, adiabatic losses from the outer shell are negligible.  The mean velocity of the shell starts at $v_{\rm ej} \approx v_{\rm ej}^{\rm max} \gtrsim 10^{4}$ km s$^{-1}$ because radiation is initially trapped, but then decreases because matter ejected later, after radiation freely escapes from the inner wind, has a lower velocity.  The energy in optical radiation is larger than that in hard radiation by a factor $\sim \tau_{\rm es}/(1-A) \gg 1$.  Hard radiation is absorbed within $\sim (1-A)^{-1}$ light crossing times, while soft radiation escapes on the longer diffusion time of $t_{\rm d} = \tau_{\rm es}t_{\rm lc}$.

\begin{figure}
\includegraphics[width=0.5\textwidth]{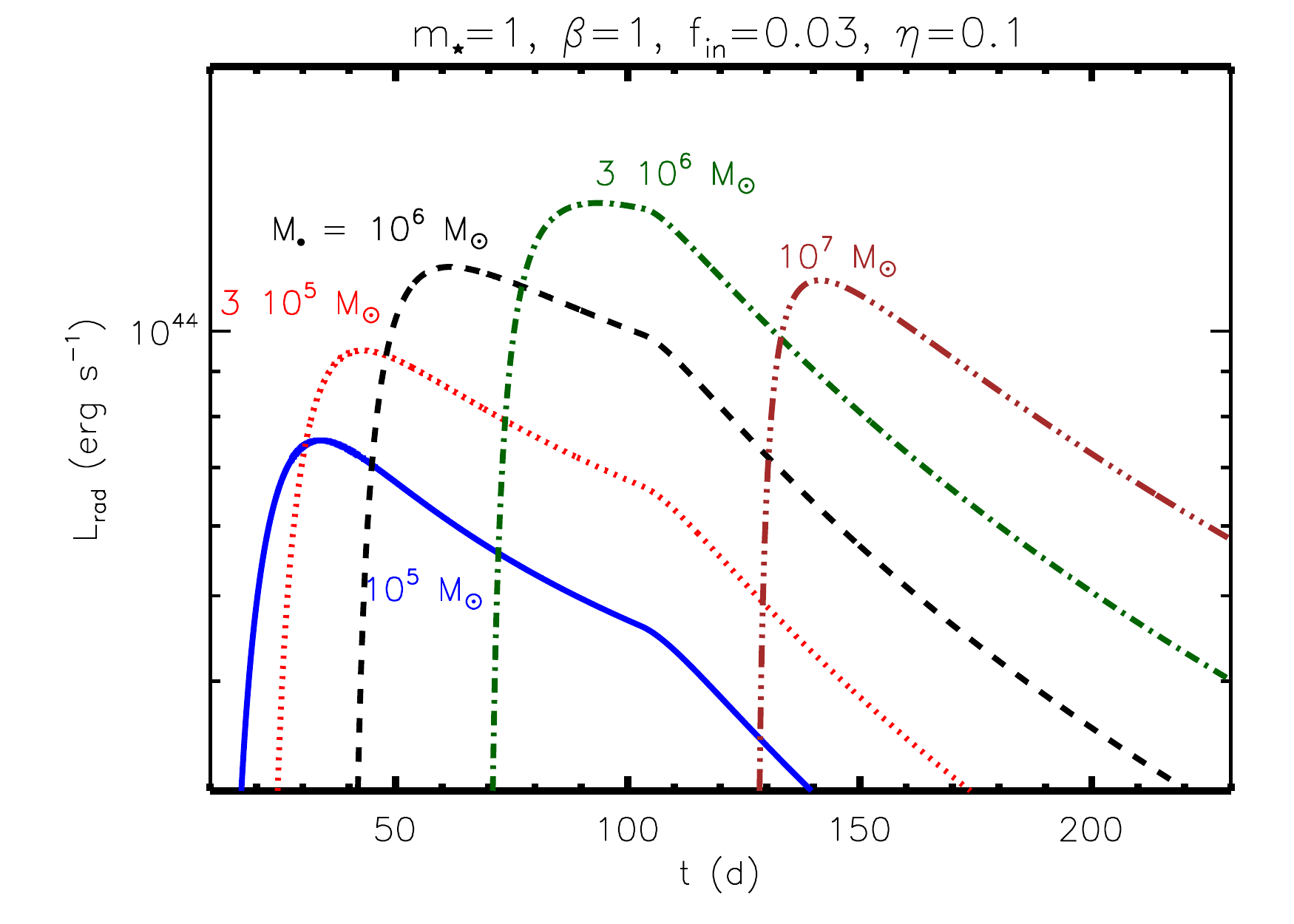}
\caption{Optical luminosity versus time since disruption, calculated for fiducial parameters $f_{\rm in} = 0.03$, $m_{\star} = 1$, $\beta = 1$, $\eta = 0.1$ and shown for different BH masses $M_{\bullet}/M_{\odot} = 10^{5}$ ({\it solid blue}), $3\times 10^{5}$  ({\it dotted red}), $10^{6}$  ({\it dashed black}), $3\times 10^{6}$  ({\it dot-dashed green}), 10$^{7}$  ({\it triple-dot dashed brown}).  Note the weak dependence of the peak luminosity on BH mass for $M_{\bullet} \lesssim M_{\bullet,\rm tr}$ (eq.~[\ref{eq:Mtr}]), as given by equation (\ref{eq:Lpk}).} 
\label{fig:LC1}
\end{figure}

\begin{figure}
\includegraphics[width=0.5\textwidth]{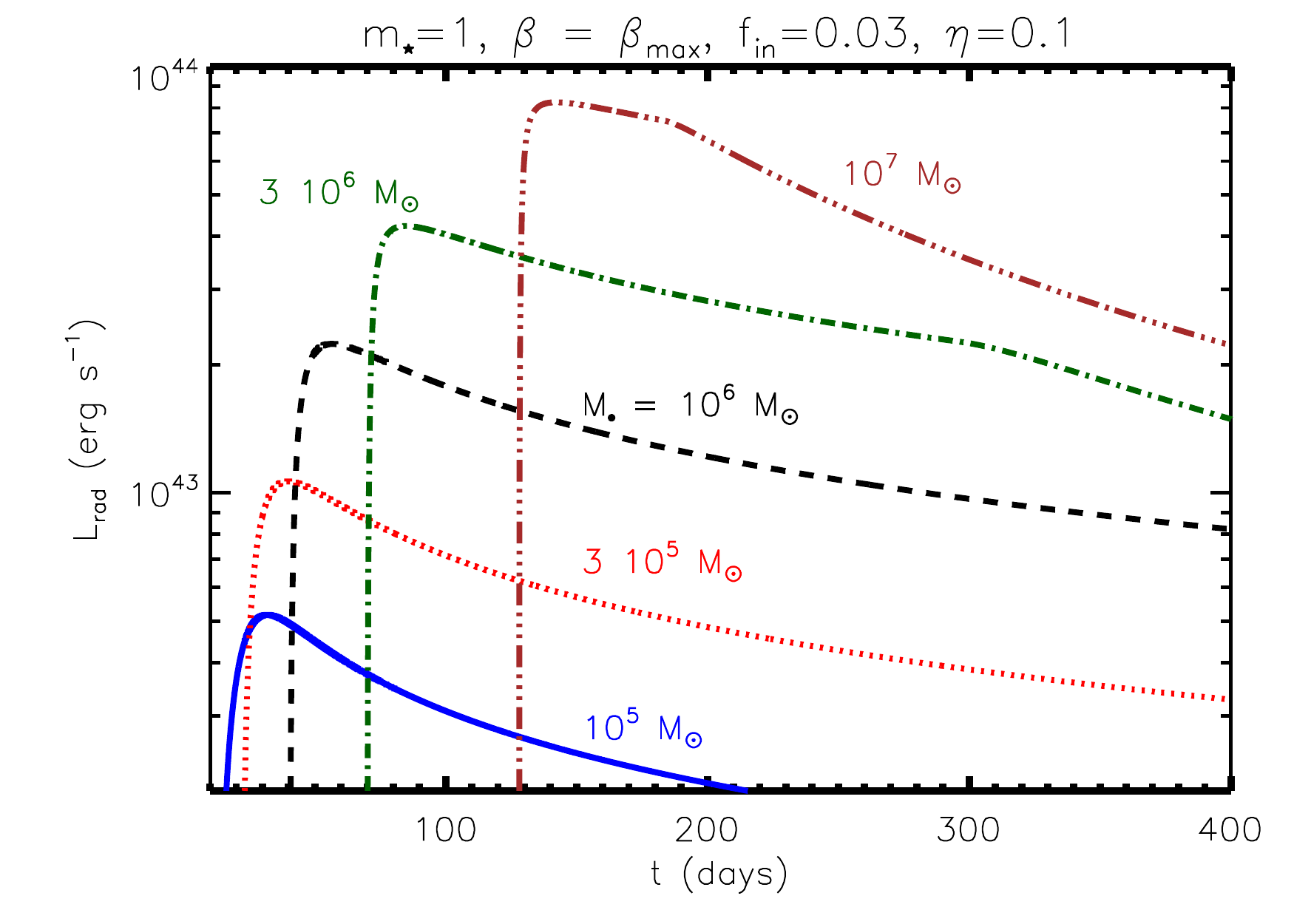}
\caption{Same as Fig.~\ref{fig:LC1}, but for $\beta = \beta_{\rm max}$ (eq.~[\ref{eq:betamax}]), i.e. for stars which are tidally disrupted on deeply plunging orbits.  Note the suppression of the peak luminosity for low mass BHs, due to adiabatic losses from the inner wind. } 
\label{fig:LC2}
\end{figure}

Figure \ref{fig:LC1} shows optical light curves for the same fiducial parameters as used for the solution in Fig.~\ref{fig:fig1}, except for a range of different BH masses, $M_{\bullet}= 10^{5}-10^{7}M_{\odot}$.  For $\beta = 1$, the optical flare typically lasts for several months with a peak luminosity of $L_{\rm pk} \gtrsim 10^{44}$ erg s$^{-1}$, relatively independent of $M_{\bullet}$.    Figure \ref{fig:LC2} is otherwise similar, but instead shows the case of a star disrupted with the largest allowable penetration factor, $\beta = \beta_{\rm max}(M_{\bullet})$, for each BH mass (eq.~[\ref{eq:betamax}]).  For a deeply penetrating disruption the emission is seen to be strongly suppressed, despite the fact that the peak fall-back rate $\dot{M}_{\rm fb}(t_{\rm fb}) \propto M_{\bullet}^{-0.5}$ is greater for small BHs, independently of $\beta$ (eq.~[\ref{eq:Mdotfb}]).  The light curve also peaks near the fall-back time, $t_{\rm fb}$, for all values of $M_{\bullet}$ and $\beta$.

\begin{figure}
\includegraphics[width=0.5\textwidth]{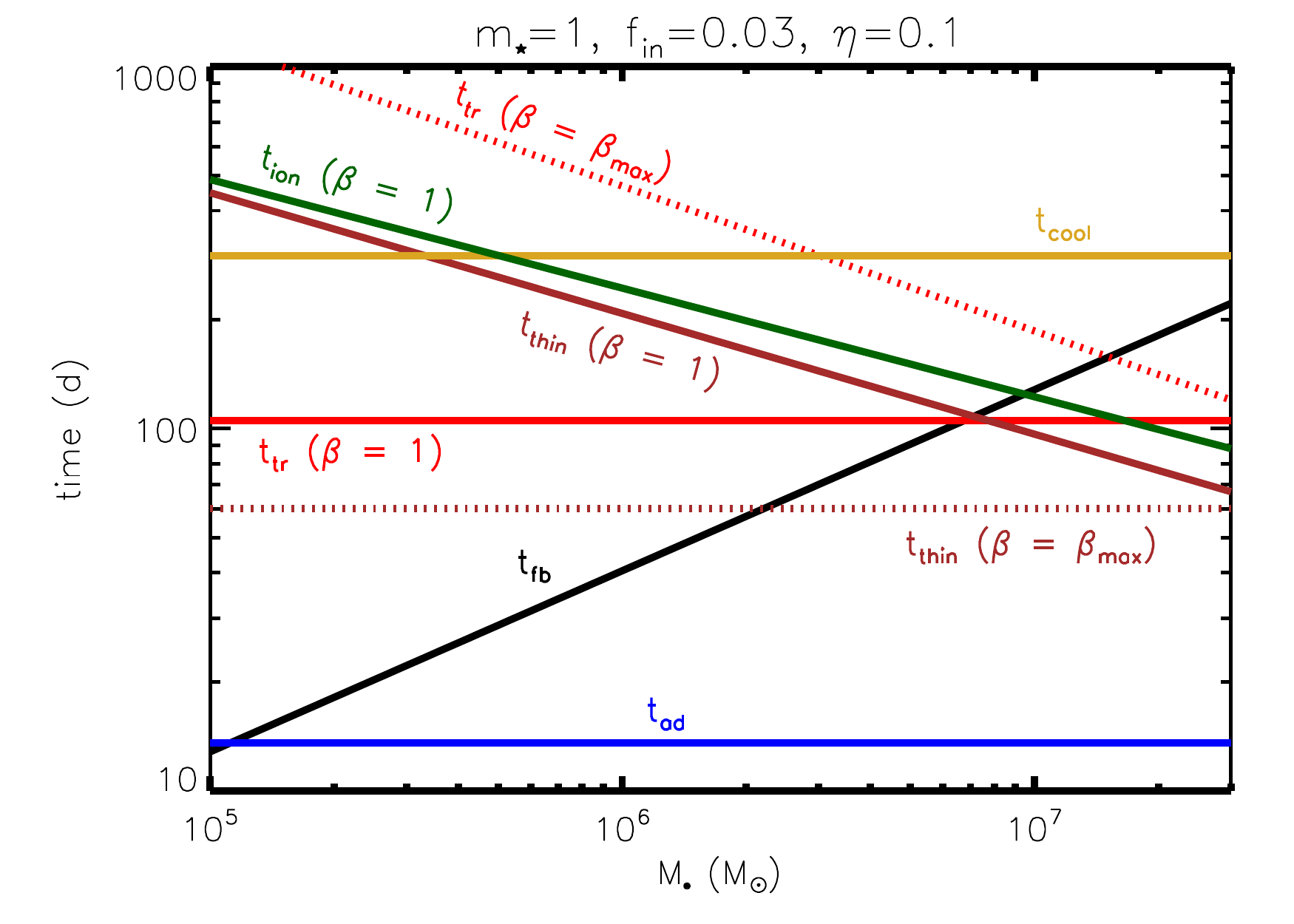}
\caption{Key timescales in days as a function of BH mass.  Black solid line: fall-back time, $t_{\rm fb} $(eq.~[\ref{eq:tfb}]), over which most mass and energy is released.  Gold solid line: radiative timescale $t_{\rm cool}$ (eq.~[\ref{eq:trad}]), prior to which matter returning to pericenter cannot cool on the dynamical time and hence powerful outflows from the disk are possible.  Blue solid line: timescale, $t_{\rm ad}$ (eq.~[\ref{eq:trapshell}]), after which radiation escapes from the outer ejecta shell without suffering adiabatic losses.  Red lines: wind trapping time $t_{\rm tr}$ (eq.~[\ref{eq:tr}]) before which optical emission is suppressed by adiabatic losses by the inner wind, shown separately for penetration factor $\beta = 1$ (solid line) and $\beta = \beta_{\rm max}$ (dashed line)  Green solid line: approximate time $t_{\rm ion}$  (eq.~[\ref{eq:tion}]) after which EUV/X-rays from the inner accretion disk ionize and escape through the entire ejecta for typical viewing angles.   Brown lines: time $t_{\rm thin}$ when the outer shell becomes optically thin to Thomson scattering (eq.~[\ref{eq:tthin}]), shown separately for penetration factor $\beta = 1$ (solid line) and $\beta = \beta_{\rm max}$ (dashed line)   } 
\label{fig:timescales}
\end{figure}

The light curves in Figs.~\ref{fig:LC1} and \ref{fig:LC2} can be readily understood from basic considerations.  To aid in the following discussion, Figure \ref{fig:timescales} shows key timescales as a function of $M_{\bullet}$, with other parameters fixed at their fiducial values of $\eta = 0.1$ and $f_{\rm in} = 0.03$.  

We begin by addressing what causes the light curves to deviate from the expected $\propto t^{-5/3}$ decay at early times.  First consider the possibility that this suppression is due to adiabatic losses from the outer shell.  Optical radiation escapes the outer shell without significant losses once the radiative loss term $L_{\rm rad}$ in equation (\ref{eq:dEoptdt2}) exceeds the adiabatic loss term, i.e. once the ratio of the diffusion to the expansion timescale,
\be t_{\rm d}/t \approx \tau_{\rm es,sh}(v_{\rm ej}/c) \approx 0.2 m_{\star}^{3/5}M_{\bullet,6}^{-1}\eta_{-1}^{-1/2}f_{\rm in,-2}^{-1/2}(t/t_{\rm fb})^{-2},
\label{eq:radefficiency}
\ee
becomes less than unity.  This condition $t_{\rm d} < t$ is satisfied after the time $t_{\rm ad}$ given by
\be
t_{\rm ad}/t_{\rm fb} \approx  0.44 \eta_{-1}^{-1/4}f_{\rm in,-2}^{-1/4} m_{\star}^{0.3}M_{\bullet,6}^{-1/2}, 
\label{eq:trapshell}
\ee
where we have assumed $v_{\rm ej} = v_{\rm ej}^{\rm max}$ (eq.~[\ref{eq:vejmax}]) because radiation is generally trapped by the inner wind on this timescale ($t_{\rm ad} \ll t_{\rm tr}$; Fig.~\ref{fig:timescales}). The fact that $t_{\rm ad} \lesssim t_{\rm fb}$ for $M_{\bullet} \gtrsim 10^{5} M_{\odot}$ implies that adiabatic losses from radiation passing through the outer shell does not suppress the peak optical luminosity for most BH masses of interest.

What about adiabatic losses from the inner wind?  As discussed earlier, the inner wind suppresses the escape of radiation for a time $t_{\rm tr}$ (eq.~[\ref{eq:ttr}]).  This trapping time is less than the fall-back time $t_{\rm fb}$ ($\sim$ timescale of peak luminosity) for BH masses above a critical value of
\be M_{\bullet,\rm tr} \equiv 6.6\times 10^{6}M_{\odot}\beta^{6/5}m_{\star}^{2/5}R_{\rm in,6}^{-6/5}.
\label{eq:Mtr}
\ee
If $M_{\bullet} > M_{\bullet,\rm tr}$ then the peak optical luminosity is simply the maximum accretion power, $L_{\rm pk} \approx L_{\rm acc}(t_{\rm fb})$.  On the other hand, if $M_{\bullet} < M_{\bullet,\rm tr}$ then $t_{\rm tr} > t_{\rm fb}$ and the peak luminosity will be reduced to $L _{\rm pk} = f_{\rm rad}L_{\rm acc}|_{t_{\rm pk}}$ and the light curve will decay after the peak time as $L_{\rm rad} \propto t^{-5/9}$ from $t = t_{\rm fb}$ until $t = t_{\rm tr}$.  At times $t \gtrsim t_{\rm tr}$ adiabatic losses from the inner wind become negligible, causing the light curve to again decay $\propto L_{\rm acc} \propto t^{-5/3}$.  This behavior matches the $M_{\bullet} \lesssim M_{\bullet,\rm tr}$ light curves shown in Figs.~\ref{fig:LC1} and \ref{fig:LC2}.  

Quantitatively, the peak optical luminosity is given by
\begin{eqnarray}
 L_{\rm pk} \approx 
\left\{
\begin{array}{lr}
   6\times 10^{43}\,{\rm erg\,s^{-1}}\beta^{-2/3}\eta_{-1}f_{\rm in,-2}M_{\bullet,6}^{0.06}m_{\star}^{0.58}R_{\rm in,6}^{2/3}, & M_{\bullet} < M_{\bullet,\rm tr}
 \\
2\times 10^{44}\,{\rm erg\,s^{-1}}\,\eta_{-1}f_{\rm in,-2}M_{\bullet,6}^{-1/2}m_{\star}^{4/5}, & M_{\bullet} > M_{\bullet,\rm tr}
\\
\end{array}
\right.
\label{eq:Lpk}
\end{eqnarray} 
where we have used equation (\ref{eq:Ld}) for $L_{\rm acc}$.  For $\beta = 1$ the peak luminosity is maximized for BH masses of $M_{\bullet} = M_{\rm \bullet,tr} \approx 6\times 10^{6}M_{\odot}$ (eq.~[\ref{eq:Mtr}]; although note the sensitive dependence on the uncertain value of $R_{\rm in}$) and $L_{\rm pk}$ depends weakly on BH mass for $M_{\bullet} < M_{\rm \bullet,tr}$ (Fig.~\ref{fig:LC1}).  However, for deeply penetrating TDEs with $\beta \sim \beta_{\rm max} \propto  M_{\bullet}^{-2/3}$ events, the peak luminosity $L_{\rm pk} \propto M_{\bullet}^{4/9}$ for $M_{\bullet} < M_{\rm \bullet,tr}$ is more highly suppressed and reaches a maximum for higher BH masses (Fig.~\ref{fig:LC2}).

Although the light curve peaks at $t \sim t_{\rm fb}$ in all cases, the bolometric output is dominated by energy released on the maximum of $t_{\rm fb}$ and $t_{\rm tr}$.  The total radiated energy is thus given by
\begin{eqnarray}
 \frac{E_{\rm rad}}{E_{\rm rad,max}} \approx
\left\{
\begin{array}{lr}
 1, & M_{\bullet} \gtrsim M_{\bullet,\rm tr}
 \\
(t_{\rm tr}/t_{\rm fb})^{-2/3}  \approx 0.6 \beta^{-2/5}M_{\bullet,6}^{1/3}m_{\star}^{-0.13}R_{\rm in,6}^{2/5}, & M_{\bullet} \lesssim M_{\bullet,\rm tr},
\\
\end{array}
\right.
\label{eq:Erad}
\end{eqnarray} 
where
\be
E_{\rm rad,max} = \eta f_{\rm in}M_{\star}c^{2}/2  \approx 10^{51}\,{\rm erg}\,\eta_{-1}f_{\rm in,-2}m_{\star}.
\ee
is the maximum radiated energy.  Values of $E_{\rm rad} \sim 10^{51}-10^{52}$ ergs for $m_{\star} \sim 1$, $\eta \sim 0.1$, and $f_{\rm in} \sim 0.01-0.1$ are consistent with the bolometric outputs of most optical TDE flares (\citealt{Gezari+08}; \citealt{Gezari+09}; \citealt{Gezari+12}; \citealt{Chornock+14}; see Fig.~2 of \citealt{Stone&Metzger15} for a compilation).

\section{Ionization State}
\label{sec:ion}

In this section we address under what conditions the ejecta will be opaque to hard nebular radiation above the ionization thresholds of hydrogen and helium.  As the density of the ejecta decreases, ionizing radiation from the central source will eventually penetrate this material and escape to infinity.  After this point, hard radiation will be thermalized much less effectively.  We consider the ionization of both the inner wind and of the outer ejecta shell, since it is not clear {\it a priori} which is the most effective at trapping the hard radiation.

Absent an external source of photo-ionization, atoms will recombine into their ground states as the ejecta expands and adiabatically cools.  For hydrogen-like species of charge $Z$, atomic mass $A = 2Z$ and mass fraction $X_A$, the bound-free opacity of neutral gas is approximately given by  (\citealt{Osterbrock&Ferland06})
\be \kappa_{\rm bf} \approx \frac{X_{A}\sigma_{\rm bf,\nu}f_{\rm n}}{2Z m_p} \approx 1.8\times 10^{6}\,{\rm cm^{2}\,g^{-1}}\,f_{\rm n}Z^{-3}\left(\frac{\nu_{\rm thr}}{\nu}\right)^{3},
\label{eq:sigmabf}
\ee
where $\sigma_{\rm bf,\nu}$ is the bound-free cross section for photons of frequency $\nu$, $\nu_{\rm thr} \approx 13.6Z^{2}$ eV is the ionization threshold frequency and $f_{\rm n}$ is the neutral fraction.  The value of $\kappa_{\rm bf}$ above the threshold is orders of magnitude higher than the electron scattering opacity, to which the ejecta already remains opaque for months to a year (eq.~[\ref{eq:taues}]).  Hard radiation therefore cannot escape to the observer unless the neutral fraction is reduced to a value $f_{\rm n} \ll 1$ by nebular photo-ionization.\footnote{Bound-free opacity was neglected in previous works on TDE outflow (e.g., \citealt{Strubbe&Quataert09}) on the grounds that Kramers opacity is subdominant compared to electron scattering opacity.  Kramers opacity, however, applies only if the atomic level populations are in collisional ionization equilibrium and does not account for photo-ionization (see Fig.~18 of \citealt{Ferland+13}).} 

The importance of photo-ionization is commonly quantified by the ionization parameter, which for the TDE ejecta is given by
\begin{eqnarray}
 \xi &=& \frac{4\pi F_{\rm H}}{n_{\rm ej}} = \frac{\pi c m_p E_{\rm H}}{n_{\rm ej}V_{\rm n}} \approx \frac{\pi m_p R_{\rm n}L_{\rm acc}}{n_{\rm ej}V_{\rm n}(1-A)} \nonumber \\
&\approx& \left\{
\begin{array}{lr}
  1.5\times 10^{3}\,{\rm erg\,cm\,s^{-1}}\,\beta^{1/2}\eta_{-1}f_{\rm in,-2}^{3/2}M_{\bullet,6}^{1/3}m_{\star}^{-7/30}(t/t_{\rm fb})^{-2/3}, & {\rm sh}
 \\
4\times 10^{3}\,{\rm erg\,cm\,s^{-1}}\,\beta^{1/2}\eta_{-1}f_{\rm in,-2}^{3/2}M_{\bullet,6}^{1/3}m_{\star}^{-7/30}, & {\rm w}
\\
\end{array}
\right.
\label{eq:xi}
\end{eqnarray} 
where $F_{\rm H} \approx E_{\rm H}c/4V_{\rm n}$ is the ionizing flux of hard radiation, $V_{\rm n} = 4\pi R_{\rm n}^{3}/3$ is volume of the nebula, and $R_{\rm n}$ is the nebula radius, which we take to equal $R_{\rm ej}$ in the case of the outer shell (``sh") (eq.~[\ref{eq:Rej}]) and $r > R_{\rm in}$ for the inner wind (``w").   In the final equality we estimate energy of ionizing radiation using equation (\ref{eq:EHapprox}).  The electron density $n_{\rm e} \approx \rho_{\rm ej}/m_p$ is calculated using equation (\ref{eq:rhoej}) in the case of the shell and equation (\ref{eq:rhow}) in the case of the wind.   For typical parameters we see that $\xi \sim 10^{3}$ erg cm s$^{-1}$, but that while $\xi$ for the wind is constant in time, $\xi$ for the outer shell decreases.  Also note that in the wind $\xi$ is to first order independent of radius, implying that the inner edge of the wind at $r \sim R_{\rm in}$ acts as the limiting step (``gate keeper") to ionizing the entire wind.  We therefore focus below on the conditions to ionize the wind on radial scales $r \approx R_{\rm in}$.

Nebular radiation will ionize the ejecta to a depth $\Delta$, set by the location at which the effective optical depth of a photon of frequency $\nu \gtrsim \nu_{\rm thr}$ to absorption reaches unity, i.e.,
\be
1 = \int_{0}^{\Delta}\rho_{\rm ej}\kappa_{\rm bf,\nu}\left[1 +\rho_{\rm ej}\kappa_{\rm es}s \right]ds \approx \tau_{\rm abs}(1 + \tau_{\rm es}),
\label{eq:tau1}
\ee
where $s$ is the depth through the ionized layer, $\tau_{\rm abs} \equiv \kappa_{\rm bf,\nu}\rho_{\rm ej}\Delta$ is the optical depth through the layer to true absorption.  The factor $1 + \tau_{\rm es}$ accounts for the additional path-length traversed by the photon due to electron scattering, where
$\tau_{\rm es} = \rho_{\rm ej}\kappa_{\rm es}\Delta$ is the electron scattering optical depth through the ionized layer.    

The neutral fraction $f_n$ is determined by the balance between photo-ionization and recombination:
\begin{eqnarray}
f_n &=& \left(1 + \frac{4\pi}{\alpha_{\rm rec} n_e}\int\frac{J_{\nu}}{h\nu}\sigma_{\rm bf,\nu}d\nu\right)^{-1}\underset{f_{n} \ll 1}\approx \frac{\alpha_{\rm rec} n_e V_{\rm n}}{c}\left(\int_{\nu_{\rm thr}}^{\infty}\frac{E_{\rm H,\nu}}{h\nu}\sigma_{\rm bf,\nu}d\nu\right)^{-1}  \nonumber \\
 &\approx& \frac{\alpha_{\rm rec} n_{e}V_{\rm n}h \nu_{\rm thr} (3-\gamma)}{\sigma_{\nu_{\rm thr}} E_{\rm H, \nu_{\rm thr}}\nu_{\rm thr} c}\approx \frac{\alpha_{\rm rec}  n_{e}V_{\rm n} h \nu_{\rm thr}(1-A) (3-\gamma)}{\sigma_{\nu_{\rm thr}} \epsilon_{\rm thr}L_{\rm acc} R_{\rm n} m_p} \nonumber \\
&\approx & 
\left\{
\begin{array}{lr}
 3\times 10^{-9}Z^{6}\frac{(1-A)(3-\gamma)}{\epsilon_{\rm thr}}T_{\rm e,4}^{-0.8}\beta^{-1/2}f_{\rm in,-2}^{-3/2}\eta_{-1}^{-1}M_{\bullet,6}^{-1/3}m_{\star}^{7/30}\left(\frac{t}{t_{\rm fb}}\right)^{2/3}, {\rm sh}
 \\
7\times 10^{-10}Z^{6}\frac{(1-A)(3-\gamma)}{\epsilon_{\rm thr}}T_{\rm e,4}^{-0.8}\beta^{-1/2}f_{\rm in,-2}^{-3/2}\eta_{-1}^{-1}M_{\bullet,6}^{-1/3}m_{\star}^{7/30},{\rm w}
\\
\end{array}
\right.
\label{eq:fn}
\end{eqnarray}
where $J_{\nu} = c E_{\rm H,\nu}/4\pi V_{\rm n}$ is the mean specific intensity of ionizing radiation, $E_{\rm H,\nu} \propto \nu^{-\gamma}$ is the nebula radiation field per unit frequency, $\alpha_{\rm rec} = 2.6\times 10^{-13}Z^{2}T_{e,4}^{-0.8}$ cm$^{3}$ s$^{-1}$ is the approximate case B radiative recombination rate, and $T_{e,4}$ is the electron temperature in the ionized layer in units of $10^{4}$ K.  We define $\epsilon_{\rm thr} \equiv  E_{\rm H, \nu_{\rm thr}}\nu_{\rm thr}/E_{\rm H} < 1$ as the fraction of the total ionizing radiation which resides at frequencies near the ionization threshold, and we have again estimated $E_{\rm H}$ using equation (\ref{eq:EHapprox}).  We expect that $\epsilon_{\rm thr} \sim 1$ because the energy of hard radiation from the accretion disk (eq.~[\ref{eq:Teffd}]) is near the He ionization threshold of 54 eV.

Using equations (\ref{eq:sigmabf}) and (\ref{eq:fn}), the ratio of bound-free to scattering opacity at the ionization threshold is given by
\begin{eqnarray}
&\zeta_{\rm thr} &\equiv \left.\frac{\kappa_{\rm bf,\nu}}{\kappa_{\rm es}}\right|_{\nu = \nu_{\rm thr}} \simeq 
\frac{X_{A}\sigma_{\rm bf,\nu_{\rm thr}}f_{\rm n}}{A m_p \kappa_{\rm es}} \nonumber \\
&\approx&  
\left\{
\begin{array}{lr}
 0.013 X_A Z^{3}\frac{(1-A)(3-\gamma)}{\epsilon_{\rm thr}}T_{\rm e,4}^{-0.8}\beta^{-1/2}f_{\rm in,-2}^{-3/2}\eta_{-1}^{-1}M_{\bullet,6}^{-1/3}m_{\star}^{7/30}\left(\frac{t}{t_{\rm fb}}\right)^{2/3}, \,\,\, {\rm sh}
 \\
3\times 10^{-3} Z^{3}X_A\frac{(1-A)(3-\gamma)}{\epsilon_{\rm thr}}T_{\rm e,4}^{-0.8}\beta^{-1/2}f_{\rm in,-2}^{-3/2}\eta_{-1}^{-1}M_{\bullet,6}^{-1/3}m_{\star}^{7/30}, \,\,\, {\rm w}
\\
\end{array}
\right.
\label{eq:zeta}
\end{eqnarray}
Photons of frequency $\nu_{\rm thr}$ penetrate to a depth $\Delta$ which can estimated from equation (\ref{eq:tau1}) to within a factor of 2 as (\citealt{Metzger+14})
\be
\Delta/R_{\rm ej} \simeq (2\tau_{\rm es})^{-1}\left[\sqrt{1 + 4 \zeta_{\rm thr}^{-1}} - 1\right] \underset{\zeta_{\rm thr} \ll 1}\approx \zeta_{\rm thr}^{-1/2}\tau_{\rm es}^{-1},
\label{eq:dth}
\ee
using the value of $\zeta_{\rm thr}$ calculated for the unattenuated nebular spectrum (eq.~[\ref{eq:zeta}]), where $R_{\rm ej}$ here represents either the thickness of the outer shell or the characteristic radial scale $\sim R_{\rm in}$ of the inner wind.  

For the disruption of a main sequence star, the ejecta is composed primarily of hydrogen (Z = 1, A = 1, $X_A \approx 0.7$) and helium (Z = 2, A = 4, $X_A \approx 0.5$), in which case equation (\ref{eq:zeta}) shows that $\zeta_{\rm thr} \lesssim 1$ for typical parameters, e.g. $\epsilon_{X} \sim 1$ and $A \approx 0.5$ (see Fig.~.3 of \citealt{Metzger+14}).  Equation (\ref{eq:dth}) then shows that the ejecta will thus remain opaque to ionizing radiation ($\Delta < R_{\rm ej}$) until the electron scattering optical depth decreases below a critical value\footnote{Equation (\ref{eq:dth}) provides the penetration depth of photons at the threshold frequency $\nu_{\rm thr}$.  Photons with frequencies above the threshold penetrate to a distance which is at most a factor of $\approx 2-3$ times larger than those at the threshold frequency (\citealt{Metzger+14}).} of
\begin{eqnarray}
&\tau_{\rm es,ion} &\approx \zeta_{\rm thr}^{-1/2}  \nonumber \\
&\approx& \left\{
\begin{array}{lr}
19 \frac{\epsilon_{\rm thr}^{1/2}}{X_A^{1/2} Z^{3/2}(3-\gamma)^{1/2}}\beta^{1/4}f_{\rm in,-2}^{3/4}\eta_{-1}^{1/2}M_{\bullet,6}^{1/6}m_{\star}^{-7/60}\left(\frac{t}{t_{\rm fb}}\right)^{-1/3}, \,\,\, {\rm sh}
 \\
40 \frac{\epsilon_{\rm thr}^{1/2}}{X_A^{1/2} Z^{3/2}(3-\gamma)^{1/2}}\beta^{1/4}f_{\rm in,-2}^{3/4}\eta_{-1}^{1/2}M_{\bullet,6}^{1/6}m_{\star}^{-7/60}, \,\,\, {\rm w}
\\
\end{array}
\right.
\label{eq:breakout}
\end{eqnarray} 
where we have taken fiducial values of $A = 0.5$ and $T_{e} = 3\times 10^{4}$ K, representative of the temperature in the photo-ionized layer ($\S\ref{sec:temperature}$).  The ejecta will be completely ionized by the inner disk, allowing hard radiation to escape, at a time
\begin{eqnarray}
\frac{t_{\rm ion}}{t_{\rm fb}} \approx
\left\{
\begin{array}{lr}
2.1 X_{A}^{0.3}Z^{0.9}(3-\gamma)^{0.3}\epsilon_{\rm thr}^{-0.3}f_{\rm in,-2}^{-1.05}\beta^{-3/4}\eta_{-1}^{-0.3}M_{\bullet,6}^{-1.1}m_{\star}^{0.71},\,\,\, {\rm sh}
 \\
4.4 X_{A}^{0.3}Z^{0.9}(3-\gamma)^{0.3}\epsilon_{\rm thr}^{-0.3}M_{\bullet,6}^{-4/5}m_{\star}^{0.41}f_{\rm in,-2}^{-3/4}\beta^{0.15}\eta_{-1}^{-0.3}R_{\rm in,6}^{-3/5} ,\,\,{\rm w}
\\
\end{array}
\right.
\label{eq:tion}
\end{eqnarray}
where we have used equations (\ref{eq:taues}) and (\ref{eq:tauw}) for $\tau_{\rm es,sh}(t)$ and $\tau_{\rm es,w}(t)$, respectively.  

Equation (\ref{eq:tion}) shows that ionizing radiation usually takes longer to tunnel through the inner wind than the outer ejecta shell.  For fiducial parameters of $M_{\bullet} \approx 10^{6}M_{\odot}$, $m_{\star} = 1$, $\beta = 1$, $\eta = 0.1$, $f_{\rm in} = 0.03$ and a spectral slope $\gamma \approx 1/3$ expected for the multi-color blackbody inner accretion disk, the ejecta should thus remain opaque to $\gtrsim 13.6$ eV radiation ($Z = 1$; $X_A$ = 0.7; $\epsilon_{\rm thr} \sim 0.1$) and to $\gtrsim 54$ eV radiation ($Z = 2$; $X_A = 0.5$; $\epsilon_{\rm thr} \approx 0.5$) for several months (Fig.~\ref{fig:timescales}).   


Equation (\ref{eq:tion}) shows that ionizing radiation will escape at earlier times as compared to a case corresponding to our fiducial parameters for (1) a larger accreted fraction $f_{\rm in}$; (2) higher radiative efficiency of the inner disk $\eta$; (3) less massive stars $m_{\star} < 1$.  Note, however, that the product of $f_{\rm in}\eta$ is constrained by the observed radiative efficiency of the flares according to equation (\ref{eq:Erad}) to near their fiducial values.  Radiation will also escape earlier than estimated by equation (\ref{eq:tion}) along directions through the ejecta with significantly lower than average density, e.g., along the axis perpendicular to the plane of the disrupted orbit.  The solid angle over which X-rays are free to escape will increase with time, approaching unity at $t \gtrsim t_{\rm ion}$.

\begin{figure}
\includegraphics[width=0.5\textwidth]{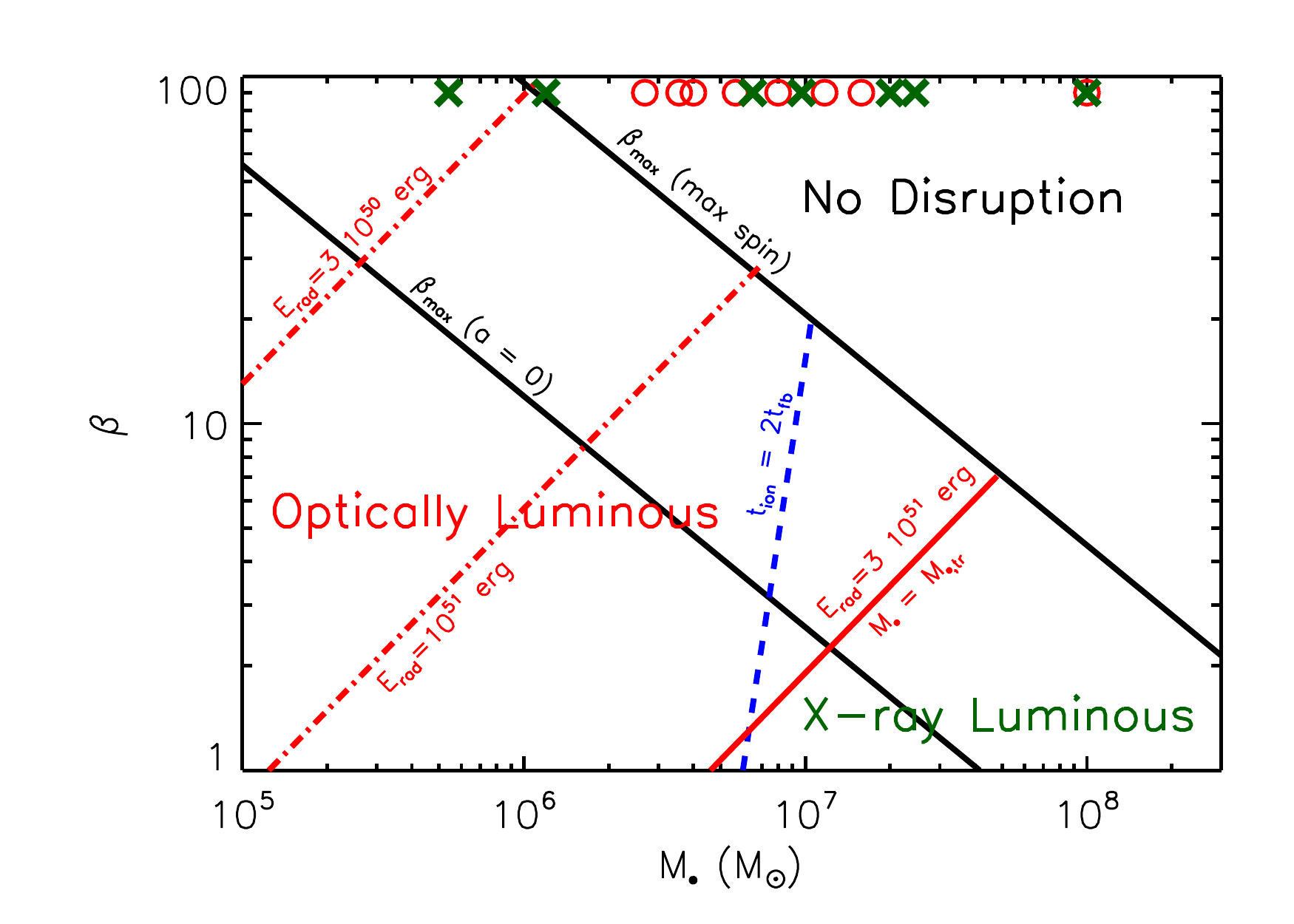}
\caption{Thermal emission from TDEs in the parameter space of BH mass $M_{\bullet}$ and penetration factor $\beta$, calculated for the disruption of a solar mass star $m_{\star} = M_{\odot}$ and fiducial parameters: $f_{\rm in} = 0.03$, $\eta = 0.1$.  For deeply penetrating orbits ($\beta > \beta_{\rm max}$; solid black line, eq.~[\ref{eq:betamax}]), little or no emission is expected because the star is swallowed whole by the BH before being tidally disrupted (shown separately for the cases of a non-spinning BH and a maximally spinning BH).  For BH masses of $M_{\bullet} \lesssim M_{\bullet,\rm tr}$ (solid red line; eq.~[\ref{eq:Mtr}]) thermal emission is suppressed because radiation is trapped by the inner wind on the fall-back time.  For the highest BH masses that allow tidal disruption, X-rays from the inner disk ionize the ejecta on timescales $\lesssim 2 t_{\rm fb}$ (dashed blue line; eq.~[\ref{eq:tion}]), allowing this emission to escape (``X-ray Luminous").  Shown with dot-dashed red lines are also the contours of constant bolometric radiated optical energy (eq.~[\ref{eq:Erad}]).  In addition to the dependence on $M_{\bullet}$ and $\beta$ highlighted by this diagram, some events may be X-ray luminous even in the ``Optically Luminous" parameter space for viewing angles with lower than average line-of-sight density, or at late times $\gg 2 t_{\rm fb}$ once the ejecta has been ionized.  Shown with circles and crosses are the estimated BH masses of those TDE flares detected in optical and/or soft X-ray frequencies, respectively (\citealt{Gezari+06,Gezari+08,Gezari+09}; \citet{Cenko+12b}, \citealt{Arcavi+14}; \citealt{Chornock+14}).} 
\label{fig:parameter}
\end{figure}

Figure \ref{fig:parameter} illustrates the parameter space of possible TDE thermal emission in the space of BH mass and penetration factor $\beta$ for fiducial parameters of $\eta = 0.1$ and $f_{\rm in} = 0.03$.  The upper right corner of the diagram corresponds to events with $\beta > \beta_{\rm max}$ (eq.~[\ref{eq:betamax}]), for which the star is swallowed whole by the BH before being disrupted.  For $\beta < \beta_{\rm max}$ there are two regimes.  For $M_{\bullet} \lesssim 10^{7}M_{\odot}$ ionizing radiation is trapped behind the inner wind for at least a couple of fall-back times (dashed blue line), reprocessing the emission to lower frequencies and producing an optically luminous flare.  Contours of total radiated emission are shown as dot-dashed lines.  For $M_{\bullet} \gtrsim 10^{7}M_{\odot}$ the ejecta becomes ionized within a couple fall-back times ($t_{\rm ion} < 2 t_{\rm fb}$; {\it dashed blue line}; eq.~[\ref{eq:tion}]), in which case much of the bolometric output should escape at  EUV/soft X-ray wavelengths, producing an X-ray luminous flare.  

In reality, the precise division between optically and X-ray luminous is theoretically uncertain given our simplistic treatment of ionization layer physics and the likely role played by viewing angle.  Shown for comparison with circles and X-rays near the top of Fig.~\ref{fig:parameter} are the inferred BH masses for the current sample of observed TDE flares, including some events detected as both optical and X-ray sources (shown with overlapping symbols).  A weak trend of higher BH masses associated with X-ray luminous TDEs is apparent, although some low mass BHs also host X-ray luminous sources, which in our model would need to result from a fortuitous viewing angle.  It should also be noted that large systematic uncertainties exist associated with BH mass measurements, which here are all taken from galaxy scaling relations.

\section{Temperature of the Optical Radiation} 
\label{sec:temperature}

Here we address the temperature of the emerging optical/UV radiation at times when the EUV/X-ray photons are absorbed by the neutral ejecta.  TDE nebulae are characterized by ionization parameters of $\xi \sim 10^{2}-10^{3}$ (eq.~[\ref{eq:xi}]), which are similar to those in other astrophysical environments.  One example is the ejecta from a classical nova, when it is irradiated by the supersoft X-rays from the heated WD surface on timescales of months to years after the nova eruption.\footnote{The velocity of nova ejecta is typically a few thousand km s$^{-1}$, similar to that of TDE outflows.  The ejected mass of $\sim 10^{-5}-10^{-4}M_{\odot}$ (and hence density) is typically a factor of $10^{4}-10^{5}$ lower than in TDEs, but the ionizing luminosity of $\sim 10^{38}-10^{39}$ erg s$^{-1}$ is typically smaller $10^{5}-10^{6}$ times smaller, resulting in a similar ionization parameter.}   Photo-ionization under such conditions heats the gas to a temperature $\sim 2-5\times 10^{4}$ K that depends only weakly on the specific radiation field due to the sensitive dependence of the rate of line cooling on temperature (see, e.g. \citealt{Cunningham+15}; their Fig.~3).  The hard radiation absorbed by the ejecta should thus be reprocessed into emission lines and thermal radiation with a characteristic energies similar to the observed temperatures of TDE flares.

If optical photons produced in the ionized layer enter equilibrium with matter as they diffused out of the ejecta, then the temperature of the radiation can be estimated as
\be
T_{\rm eff}^{\rm TE} = \left(\frac{L_{\rm rad}}{4\pi \sigma R_{\rm ej}^{2}}\right)^{1/4} \approx 2.1\times 10^{4}\,{\rm K}\,\,\eta_{-1}^{1/4}f_{\rm in,-2}^{1/4}\beta^{-1/4}m_{\star}^{13/60}M_{\bullet,6}^{-13/24}\left(\frac{t}{t_{\rm fb}}\right)^{-11/12}, 
\label{eq:Teff}
\ee
where we have used equation (\ref{eq:Rej}) for $R_{\rm ej}$ and in the second line we have assumed $t > t_{\rm tr}$ (eq.~[\ref{eq:tr}]), i.e. that the radiated luminosity tracks the accretion power, $L_{\rm rad} = L_{\rm acc} \propto t^{-5/3}$.  

Equation (\ref{eq:Teff}) predicts characteristic temperatures within a factor of a few of those of observed optical TDE flares.  However, it also predicts that the temperature should decline in time, in contradiction with the temporally constant temperatures of optically-discovered TDE flares.  TDE spectra, unlike those of most supernovae, do not evolve to become redder with time as the ejecta expands and cools (e.g., \citealt{Chornock+14}).

The true color temperature of the emission will, however, remain higher than that predicted by equation (\ref{eq:Teff}) because as the ejecta spreads out and becomes optically thin, the photosphere will recede deeper than our naive estimate of $R_{\rm ej}$ used above.  Note in particular that for $M_{\bullet} \gtrsim 10^{6}M_{\odot}$, the outer ejecta shell becomes optically Thomson-thin within less than a few fall-back times (Fig.~\ref{fig:timescales}).  The true photosphere radius will thus increase with time less rapidly than $\propto t$ (and may even decrease), resulting in a much slower evolution of $T_{\rm eff}$.  Due to the lower scattering optical depth of the outer layers, the mean photon energy of $\sim$ few 10$^{4}$ K set by the temperature of the inner photo-ionized ejecta may well be preserved in the emission that escapes to the observer, even though the ejecta is expanding as a function of time.  Future work is required to more accurately estimate the effective temperature of the reprocessed emission.

\section{Implications for Cosmic Evolution of Supermassive Black Hole Spin}
\label{sec:spin}

Our model has significant implications for the mass and spin evolution of low-mass BHs.  Theoretically (observationally) calculated TDE rates $\dot{N}\sim 10^{-4} ~(10^{-5})~{\rm gal}^{-1}~{\rm yr}^{-1}$ (cf.~\citealt{Stone&Metzger15}, \citealt{VanVelzen+11}) imply that BHs with $M_\bullet \lesssim 10^6 M_\odot ~(10^5 M_\odot)$ should see significant growth from star capture over a Hubble time, and may even place a lower limit on BH mass, {\it provided} the traditional assumption that $f_{\rm in} \sim 1$ is true.  

\begin{figure}
\includegraphics[width=0.5\textwidth]{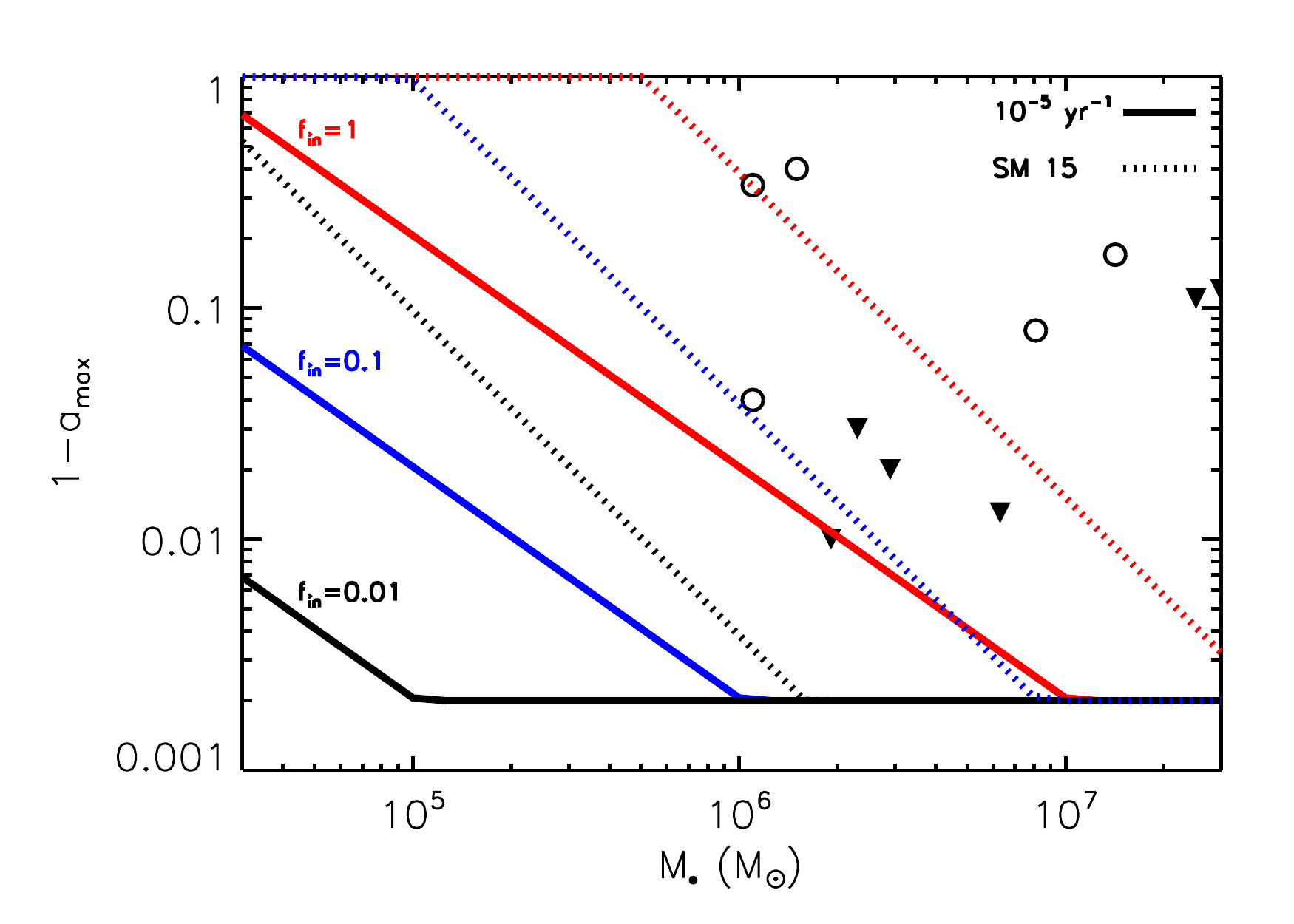}
\caption{The maximum attainable supermassive BH spin, $a_{\rm max}$, for BHs of a given mass $M_\bullet$ which have grown partially through stellar tidal disruption.  We plot $1-a_{\rm max}$ to show the variation in very high values of BH spin.  Red, blue, and black lines assume $f_{\rm in}=1$, $0.1$, and $0.01$, respectively.  The solid lines show an observationally estimated TDE rate of $\dot{N}=10^{-5}~{\rm gal}^{-1}~{\rm yr}^{-1}$ \citep{vanVelzen&Farrar14}, while the dashed lines show the (higher) theoretically predicted TDE rate of $\dot{N} = 2.9 \times 10^{-5}~{\rm gal}^{-1}~{\rm yr}^{-1}~(M_\bullet / (10^8M_\odot))^{-0.404}$ from \citet{Stone&Metzger15}.  All curves asymptote to the Thorne limit ($a_{\rm max}=0.998$) once the BH mass becomes large enough that star capture is a negligible contribution to the BH's spin evolution.  In the traditional picture, where $f_{\rm in}=1$, TDEs place a more restrictive cap on BH spin than does accretion physics for $M_\bullet \lesssim 10^7 M_\odot$, but this is not the case in our revised TDE model with $f_{\rm in} \ll 1$.  Shown with circles and arrows are published measurements and upper limits, respectively, on BH spin, compiled by \citet[Table 1]{Reynolds13}. We assume the accretion of stars with an average mass $M_\star = 0.3 M_\odot$.}
\label{fig:amax}
\end{figure}

In our model, $f_{\rm in} \ll 1$ and TDEs are unable to contribute to the growth of BHs in typical galaxies.  In the traditional picture ($f_{\rm in}= 1$), TDEs can contribute to spin evolution even at larger ($\sim 10^7 M_\odot$) BH masses.  The Thorne limit of $a_\bullet = 0.998$ \citep{Thorne74} is often taken to represent the maximum spin of BHs in the universe\footnote{However, the maximum spin that can be reached through constant prograde accretion depends on accretion mode, and the Thorne limit strictly applies only to thin disks.  Most optimistically, \citet{Sadowski+11} find that super-Eddington and low-viscosity slim disk accretion can saturate at 
$a_{\rm max}=0.9995$; see references therein for summaries of other models that find smaller saturation spins.}, and BHs that grow through the accretion of coplanar gas in thin disks will eventually reach it \citep{Berti&Volonteri08}.  However, because the set of tidally disrupted stars has little net angular momentum, their net effect over time is to despin rapidly rotating BHs, as shown in Figure \ref{fig:amax}.  If TDEs result in the accretion of a mass $M_\star/2$ of gas onto the BH, then even a relatively low TDE rate can impose a different spin limit on BHs: 
\begin{equation}
a_{\rm max} \approx 1-\dot{N}f_{\rm in}t_{\rm H}M_\star/(2M_\bullet), \label{eq:aMax}
\end{equation}
where we have assumed isotropic final orbits for the disrupted stars.  A more accurate estimate of $a_{\rm max}$ can be achieved using Eqs. 5.267 of \citet{Merritt13}, which describe exactly the relativistic, inclination-dependent change in BH spin following accretion of stellar matter; however, the simple Newtonian estimate of equation (\ref{eq:aMax}) is sufficient to highlight the large role TDEs can play in the spin evolution of smaller supermassive BHs.

Even low, observationally inferred, TDE rates are capable of spinning $10^5 M_\odot$ ($10^6 M_\odot$) BHs down to an $a_{\rm max} \sim 0.8 ~(0.98)$ in the traditional picture of tidal disruption.  Our model, however, predicts that the TDE contribution to BH spin evolution is vastly weaker at all masses, though not totally negligible.  

Strictly speaking, our assumption of isotropic final orbits for tidally disrupted stars is only valid for TDEs sourced from two-body relaxation in a spherical, non-rotating cluster; in this limit, a BH that grew only through the capture of $N$ stars would statistically approach $\langle a_\bullet \rangle \approx 1/\sqrt{N}$.   If the stellar cusp possesses net rotation, tidally disrupted stars will also, and Eq. \ref{eq:aMax} would be modified to $a_{\rm max} \approx 1-\dot{N}(1-f_{\rm pin})\chi f_{\rm in}t_{\rm H}M_\star/(2M_\bullet) - \dot{N}f_{\rm pin} f_{\rm in}t_{\rm H}M_\star/(2M_\bullet)$, where $\chi$ is the fraction of stars on retrograde orbits and $f_{\rm pin}$ is the fraction of TDEs from the ``pinhole'' regime of disruption\footnote{In the pinhole regime of tidal disruption, the per-orbit change in specific angular momentum $|\Delta J| \gg |J_{\rm LC}|\equiv\sqrt{2GM_\bullet R_{\rm t}}$, the specific angular momentum of the loss cone.  In this regime the star wanders in and out of the loss cone many times per orbit, and its final, disruptive passage through the loss cone is just as likely to be prograde as it is to be retrograde with respect to any reference axis.  In the opposite, ``diffusive,'' regime, $|\Delta J| \ll |J_{\rm LC}|$ and disrupted stars retain a memory of the cluster rotation axis.}; because $f_{\rm pin} \sim 1$ when $M_\bullet \lesssim 10^7 M_\odot$ \citep{Stone&Metzger15}, we neglect the effects of stellar cusp rotation\footnote{\citet{Beloborodov+92} examined the equilibrium spins of supermassive BHs that grew through tidal disruption, but assuming that the tidally liberated gas only accreted after settling into the cluster equatorial plane.  In our terminology, this is equivalent to the extreme case $f_{\rm pin}=0$ and $\chi=0$.}.  Our isotropic approximation is further justified by the fact that two-body relaxation may not be the dominant dynamical process for creating TDEs.  TDE rates can be enhanced above the level set by two-body relaxation if axisymmetry \citep{Magorrian&Tremaine99, Vasiliev&Merritt13} or triaxiality \citep{Merritt&Poon04} exist in the stellar potential; in this case, orbits will not conserve all components of angular momentum, and some will collisionlessly wander to very small pericenter.  These aspherical channels for producing TDEs primarily funnel stars to the BH in the ``pinhole'' regime of disruption \citep{Vasiliev&Merritt13}, which severely reduces the net spin they can impart to the BH. 


In practice, TDEs have the greatest impact on $a_{\rm max}$ if BHs grow through ``coherent'' gas accretion, i.e. accretion of gas from a fixed orbital plane.  Whether or not this occurs depends on the larger-scale physics of the host galaxy.  State of the art cosmological simulations (with subgrid models for BH spin evolution) find that the small BHs which likely dominate TDE rates ($10^6 < M_\bullet / M_\odot < 10^7$) spin up efficiently through gas accretion, generally coming close to $a_\bullet=0.998$ \citep[Figure 7]{Dubois+14}.  As shown in our Figure \ref{fig:amax}, this mass range is one in which TDEs render the Thorne limit unattainable if $f_{\rm in} \approx 1$, and the true $a_{\rm max}$ at $10^6 M_\odot$ is between $0.6$ and $0.98$, depending on the TDE rate.  Roughly 20 supermassive BH spins have been measured using the Fe K$\alpha$ technique, and tantalizingly, several fall into the region of parameter space that can constrain $f_{\rm in}$.  MCG-6-30-15 has $M_\bullet =2.9^{+1.8}_{-1.6} \times 10^6 M_\odot$ and $a_\bullet > 0.98$, Ark564 has $M_\bullet \sim 1.1 \times 10^6M_\odot$ and $a_\bullet = 0.96^{+0.01}_{-0.11}$, IRAS13224-3809 has $M_\bullet \sim 6.3 \times 10^6$ and $a_\bullet > 0.987$, NGC 4051 has $M_\bullet =1.91\pm 0.78 \times 10^6 M_\odot$ and $a_\bullet > 0.99$, and 1H0707-495 has $M_\bullet \sim 2.3 \times 10^6M_\odot$ and $a_\bullet > 0.97$ \citep[Table 1]{Reynolds13}.  

All of these mass-spin combinations are impossible if $f_{\rm in}=1$ and their host galaxies have TDE rates in line with the predictions of \citet{Stone&Metzger15}.  Several are even more constraining; for example, the measured mass-spin pair in NGC 4051 cannot be produced by a galaxy following our theoretically predicted TDE rates unless $f_{\rm in} \lesssim 0.1$.  Although this comparison with data is suggestive, we caution that $f_{\rm in}$ can only be rigorously constrained through statistical analysis of a large sample of measured supermassive BH spins, a task beyond the immediate scope of this paper.

\section{Discussion and Conclusions}
\label{sec:discussion}

We have presented a model for the thermal optical emission from TDEs, with implications for a possible unification of thermal TDE phenomenon.  Our main assumption is that only a small fraction $f_{\rm in} \sim 0.01-0.1$ of the initially bound stellar debris ends up reaching the inner circular disk and accreting onto the BH, with the remainder (comprising roughly half of the original star) blown out in a wind or unbound stream (\citealt{Strubbe&Quataert09}; \citealt{Lodato&Rossi11}) that rapidly comes to encase the BH.  

Such a low accretion efficiency is suggested by numerical simulations of adiabatic accretion with low angular momentum and negligible initial binding energy (\citealt{Li+13}), which find that only a fraction $\sim \alpha$ of the inflowing gas is accreted by the central BH.  The precise value of the effective viscosity $\alpha$ depends on the saturation level of MHD turbulence, which could well be different for shear in the eccentric gaseous stream of a TDE than in a circular accretion disk (\citealt{Guillochon&RamirezRuiz15}).  The efficacy of MHD circularization has been discounted as the primary method of disk circularization on the grounds that $\sim 1/\alpha \gg 1$ orbital periods are required to fully circularize the material (\citealt{Guillochon&RamirezRuiz15}).  However, here we have emphasized that the energy which must be released to circularize even a small fraction $\alpha \gtrsim 0.01$ of the inflowing gas is sufficient to unbind the remaining fraction $1-\alpha$, given the weak binding energy of the debris (eq.~[\ref{eq:Eratio}]) and its effectively adiabatic evolution due to radiation trapping.    Numerical simulations of super-Eddington accretion (e.g., \citealt{Jiang+14}) are not directly applicable to the problem of TDE fall-back if they initialize a disk which is already tightly bound to the BH.  

The importance of unbound outflows in radiatively inefficient accretion flows remains an issue of debate (\citealt{Yuan&Narayan14}), so the assumption of low $f_{\rm in}$ must be taken as an assumption of our model rather than a proven fact.  However, even if only a few percent of the star accretes, we find that the radiative efficiency of reprocessing by the outflowing debris is high enough to explain the bolometric output of most TDE flares (eq.~[\ref{eq:Erad}]).  This challenges the implicit assumption of recent work that the majority of the initially bound debris must indeed circularize to explain the observed thermal flares.  From this viewpoint, a larger fraction of the star would accrete if the stellar debris were more tightly bound.  This could occur if a sub-population of TDEs arise from very tightly bound stellar orbits \citep{Hayasaki+13}, or if the interaction between misaligned BH spin and tidal compression in a high $\beta$ event could dramatically increase $E_{\rm t}$ \citep{Stone+13}.  However, both of these possibilities are speculative and unlikely to occur except in a small minority of TDEs.  

The small amount of energy released in binding the fraction $f_{\rm in}$ to the disk implies relatively low velocities of a few $10^{3}$ km s$^{-1}$ for the bulk of the material, although at early times matter can be accelerated by the pressure of trapped radiation to a higher velocity of $\gtrsim 10^{4}$ km s$^{-1}$ (eq.~[\ref{eq:vejmax}]).  These velocities are broadly consistent with the widths of the emission lines observed in the TDE optical spectra of $\sim 3-10\times 10^{4}$ km s$^{-1}$ (\citealt{Gezari+12}; \citealt{Arcavi+14}).  These lines may originate directly from the inner X-ray ionized layer of the ejecta, or from the faster outflow expected on larger scales.  More detailed photo-ionization and radiative transfer calculations will be required to create more concrete predictions for these emission line diagnostics, and how they vary with the other properties of the system using considerations similar to those in \citet{Guillochon+14}.

One of the biggest mysteries associated with the current observational sample of thermal TDE flares is what distinguishes the optically and X-ray luminous sources.  For BHs of mass $M_{\bullet}$ above a critical threshold $10^{7}M_{\odot}$ (which depends on $f_{\rm in}$, $\eta$, $R_{\rm in}$, and $m_{\star}$), we find that ionizing radiation from the inner disk is trapped behind the wind ejecta for timescales of several months or longer, similar to the characteristic duration of TDE light curves.  The irradiated inner wind on a radial scale of $\gtrsim 10R_{\rm p}$ can thus provide the ``reprocessing" layer needed to produce bright optical emission, without requiring a radially-extended distribution of mass due to inefficient circularization.  

Accretion power is converted into reprocessed radiation with high efficiency because ionizing radiation is absorbed by the neutral ejecta on at most a few light crossing times across the nebula.  The diffusion timescale of the reprocessed optical radiation by the outer ejecta is generally short compared to the expansion timescale for BH masses of $M_{\bullet} \gtrsim 10^{5}M_{\odot}$ (eq.~[\ref{eq:trapshell}]; Fig.~\ref{fig:timescales}).  However, for $M_{\bullet} \lesssim M_{\bullet,\rm tr} \sim$ few $10^{6}M_{\odot}$ (eq.~[\ref{eq:Mtr}]), trapping by the inner wind reduces the radiative efficiency at early times.  In this case, there is an early phase where the radiated luminosity $L_{\rm rad} \propto t^{-5/9}$ before the light curve approaches the the well-known $\propto t^{-5/3}$ decay (Fig.~\ref{fig:LC1}, \ref{fig:LC2}).  The bolometric output of the TDE flare can therefore be considerably suppressed for low-$M_{\bullet}$ and/or high $\beta$ events.  The temperature of the emission may be regulated to a relatively constant value of several $10^{4}$ K by the balance between photo-ionization and line cooling, similar to X-ray irradiation of nova ejecta of similar ionization parameter (eq.~[\ref{eq:xi}]).  


The model developed here resembles the disk wind models of \citet{Strubbe&Quataert09} and \citet{Lodato&Rossi11}, which have been discounted as the source of observed optical emission based on both spectral line widths and a predicted fast light curve decay ($\propto t^{-95/36}$), in disagreement with observations (e.g.~\citealt{Gezari+12}).  We find that these deficiencies are remedied by considering higher ejecta masses and slower outflow velocities than in past work (which are also more consistent with the findings of numerical simulations and basic energetic considerations), and by taking into account the role of bound-free opacity on the escape of ionizing radiation.  Our model is similar to that which has been developed for optical emission from stellar explosions powered by a central engine, which is one of the most promising models for super-luminous supernovae (SLSNe; \citealt{Kasen&Bildsten10}; \citealt{Woosley10}; \citealt{Metzger+14}).  It is interesting to note that the blue colors and spectra of some TDE flares qualitatively resemble those of SLSNe (\citealt{Chornock+14}).

For more massive BHs with $M_{\bullet} \gtrsim $ 10$^{7}M_{\odot}$ (the precise value depending on $f_{\rm in}$ and $m_{\star}$), X-rays from the inner disk may completely ionize and escape the ejecta on a timescale $t_{\rm ion}$ (eq.~[\ref{eq:tion}]) exceeding a few fall-back times of the most bound debris.  If and once this occurs, reprocessing to optical frequencies will be less effective and the event will be more X-ray luminous.  Such cases may explain the thermal X-ray detected flares (e.g., \citealt{Bade+96}, \citealt{Grupe+99}, \citealt{Komossa&Bade99}, \citealt{Komossa&Greiner99}, \citealt{Donley+02}, \citealt{Esquej+08}, \citealt{Maksym+10}, \citealt{Saxton+12}).  

Even in optically luminous flares where no X-ray emission is initially detected, X-rays may escape at late times as the density of the expanding shell decreases and the ionization front reaches the surface.  We strongly encourage late-time X-ray monitoring of optically-discovered TDEs on timescales of months or years after the disruption, to search for this ``ionization break-out" signal.  The prediction that X-ray luminous TDEs may preferentially occur in higher mass galaxies appears to be in tension with observations of X-ray flares in some low mass galaxies (Fig.~\ref{fig:parameter}), but it is hard to draw firm conclusions due to both small-number statistics and the unreliable BH mass estimates in current TDE candidates. 

Even if X-rays are trapped by the bulk of the ejecta, escape is possible along directions with lower than average density ($t_{\rm ion} \propto n_{e}^{0.3}$; eq.~[\ref{eq:tion}]).  These viewing angles may include directions perpendicular to the orbital plane of the disrupted star, or opposite the dominant direction of the outflowing gas if the circularizing debris cannot phase mix before most of the mass is unbound.  TDEs viewed along such low-density `holes' could be both X-ray and optically-luminous.  Even if our assumption of quasi-spherical ejecta is not justified and ionizing radiation is not trapped behind the neutral outflow for more than a light crossing time, those X-rays that do intercept dense streams of ejecta should thermalize with high efficiency, given the low albedo $A \sim 0.4-0.7$ which is expected to characterize the exposed inner edge of the ejecta for  the ratio of absorption to scattering opacities of relevance (compare eq.~[\ref{eq:zeta}] with Fig.~3 of \citealt{Metzger+14}).     

\citet{Stone&Metzger15} highlight the tension, of at least one order of magnitude, between the rates of TDEs predicted by modeling two-body scattering of stars in galactic nuclei of $\sim 10^{-4}$ gal$^{-1}$ yr$^{-1}$ (see also \citealt{Wang&Merritt04}, \citealt{Merritt15}), and the detected rates of optically-luminous TDEs of $\sim 10^{-5}$ gal$^{-1}$ yr$^{-1}$ (e.g.~\citealt{VanVelzen+11}).  Part of this discrepancy could be the result of dust obscuration in galactic nuclei or observational selection effects; however, part of it could be attributed to only a small fraction of all TDEs being optically luminous.  
Our finding that the peak luminosities and optical fluences of TDE flares are smaller for low mass BHs than would be naively anticipated from their peak accretion rate (eq.~[\ref{eq:Lpk}]) would act to suppress the rate of TDE detection in low mass galaxies.  Indeed, the inferred BH masses of optically-discovered TDE flares seem to cluster around $M_{\bullet} \sim 10^{7}M_{\odot}$, despite the theoretical expectation that the detected TDE rate by a flux-limited survey should be dominated by lower-mass BHs, assuming peak optical luminosity tracks the mass fall-back rate alone \citep{Stone&Metzger15}. 



Our finding that most TDEs may accrete only a small fraction of their debris makes the jetted class of TDEs all the more mysterious.  The prototypical jetted TDE, {\it Swift} J1644+57, radiated $\sim 10^{54}$ ergs of isotropic X-rays (\citealt{Bloom+11}; \citealt{Burrows+11}), while the kinetic energy of the jet responsible for the radio afterglow likewise approached a significant fraction of a solar rest mass $\sim 10^{52}-10^{53}$ ergs (e.g.,~\citealt{Berger+12}; \citealt{BarniolDuran&Piran13}; \citealt{Tchekhovskoy+14}; \citealt{Mimica+15}).  Explaining this required accretion power might require a massive star  in the context of low accretion efficiency $f_{\rm in} \sim 0.01-0.1$, or some unusual circumstances, e.g.~more tightly bound stellar debris due perhaps to an eccentric stellar trajectory or effects of BH spin.  Interestingly, one of the consequences of our assumption of a low accreted fraction is that the true Eddington ratio of the accretion rate reaching the BH is a factor of $\sim 10-100$ times lower than that of the fall-back (eq.~[\ref{eq:edd}]).  In this case, rather low BH masses of $\ll 10^{6}M_{\odot}$ may be required to produce highly super-Eddington accretion on smaller scales.  We speculate that if highly super-Eddington accretion is the key requirement to producing a powerful jet, that this condition might only be satisfied by potentially rare low mass BHs.  
 


The potentially out-sized role of TDEs in supermassive BH spin evolution, which to our knowledge has not been examined before, turns out to depend critically on the value of $f_{\rm in}$.  For traditional values of $f_{\rm in} \sim 1$, it is not possible for small supermassive BHs to reach the theoretical upper limits on $a_\bullet$ predicted by \citet{Thorne74} and successor works.  Our prediction, that $f_{\rm in} \ll 1$, gives TDEs much less leverage to cap supermassive BH spin.  The Fe K$\alpha$ measurement of several rapidly spinning BHs with $M_\bullet \sim 10^6 M_\odot$ is hard to reconcile with theoretically predicted TDE rates \citep{Stone&Metzger15} and the assumption that $f_{\rm in}=1$.  Future analyses of Fe K$\alpha$ spin samples that account for systematic errors and small number statistics should be able to rigorously constrain a combination of the true TDE rate and $f_{\rm in}$.

Our model will be improved in future work in several directions.  First, we will employ a one or multi-dimensional outflow model to replace our simplified one-zone approach.  Also needed is a self-consistent photo-ionization calculation to more accurately determine the conditions for ionizing radiation to escape, and to explore emission line diagnostics.  

\section*{Acknowledgments}

We acknowledge helpful conversations with Aleksey Generozov, David Merritt, Jerry Ostriker, Sjoert Van Velzen and Indrek Vurm.  BDM
gratefully acknowledges support from NASA {\it Fermi} grant
NNX14AQ68G, NSF grant AST-1410950, and the Alfred P. Sloan Foundation.

\bibliographystyle{mn2e}
\bibliography{ms}


\end{document}